\documentclass[reprint,pre,superscriptaddress]{revtex4-1}
\usepackage{blindtext}
\usepackage{amsmath, amssymb, graphicx, mathtools, multirow,makecell,subfigure,color,float,stmaryrd,mathrsfs}

\usepackage{enumitem}

\usepackage[utf8]{inputenc}

\usepackage{ulem}

\makeatletter
\renewcommand{\p@subsection}{}
\renewcommand{\p@subsubsection}{}
\makeatother

\makeatletter
\newsavebox{\@brx}
\newcommand{\llangle}[1][]{\savebox{\@brx}{\(\m@th{#1\langle}\)}%
  \mathopen{\copy\@brx\mkern2mu\kern-0.9\wd\@brx\usebox{\@brx}}}
\newcommand{\rrangle}[1][]{\savebox{\@brx}{\(\m@th{#1\rangle}\)}%
  \mathclose{\copy\@brx\mkern2mu\kern-0.9\wd\@brx\usebox{\@brx}}}
\makeatother

\newcommand{\norm}[1]{\left\lVert#1\right\rVert}

\usepackage{hyperref}
\hypersetup{
	colorlinks=true,       
	linkcolor=red,          
	citecolor=blue,        
	filecolor=magenta,      
	urlcolor=blue           
}

\begin{document}
	
	\title{Active drive towards  elastic  spinodals }

	\author{Ayan Roychowdhury}
	\affiliation{Simons Centre for the Study of Living Machines, National Centre for Biological Sciences-TIFR, Bengaluru, India  560065.}
	\affiliation{PMMH, CNRS–UMR 7636, PSL-ESPCI, 10 Rue Vauquelin, Paris 75005, France.}
	\author{Madan Rao}
	\affiliation{Simons Centre for the Study of Living Machines, National Centre for Biological Sciences-TIFR, Bengaluru, India  560065.}
	
	\author{Lev Truskinovsky}
	\affiliation{PMMH, CNRS–UMR 7636, PSL-ESPCI, 10 Rue Vauquelin, Paris 75005, France.}

	\begin{abstract} 
 	Active matter, 
    exemplified by adaptive living materials such as the actomyosin cytoskeleton, can navigate material parameter space dynamically, leading to unconventional mechanical responses.
    In particular, it can self-drive toward {\it elastic spinodal} regimes, where inhomogeneous floppy modes induce elastic degeneracy and enable a controlled interplay between rigidity loss and recovery. Proximity to such marginal states leads to stress localization and the formation of force chains that can be actively assembled and disassembled. 
    Here, we extend the classical notion of spinodal states to active solids and demonstrate how these extreme mechanical regimes can be actively accessed.  Moreover, we show that in a nonlinear setting, crossing elastic spinodals generates new energy wells and  makes force channeling  an
    intrinsic feature of the emerging microstructure. 
	\end{abstract}

	\maketitle
	
	\section{Introduction}
	
	In passive systems elastic  rigidity   usually emerges   as a result of  breaking continuous symmetry  \cite{ChaikinLubenskybook}. It can also appear  in the process of introducing overconstraining   interactions
\cite{Maxwell1864,Ellenbroeketal2009,EdwardsGrinev1999,LiuNagel1998,Oncketal2005}  
or  as a result of   tuning   of the    pre-stress \cite{Shlomo1998,  Ellenbroeketal2009,Ingberetal2014,LeeMerkel2022,MaoLubensky2018,Merkeletal2019,Calladine1978}. Similarly, it is known that  rigidity can be   lost  in passive systems  due to  symmetry restoration \cite{HalperinNelson1978, ChaikinLubenskybook},
through  underconstraining  
 \cite{Goodrichetal2015,catesetal1998a} 
and  by   relaxing  the  pre-stress \cite{ChaikinLubenskybook, Goodrichetal2015,Shlomo1998, Roncerayetal2016,Pritchard2014}.

Living materials,  operating far from  equilibrium, can  be more flexible in manipulating the partial  loss and  recovery of elastic rigidity \cite{ChaikinLubenskybook, Goodrichetal2015,Shlomo1998}.  This process is often driven  by the transduction of metabolic resources into functional work
\cite{HawkinsLiverpool2014,HumphreyKas2002,Koenderinketal2009,MunroGardel2021,Oriolaetal2017,Weirichetal2017,Xietal2019,HiraiwaSalbreux2016, 
 Sheshkaetal2016, maitravoituriez2020, beheraetal2023, Schillersetal2010,  Etienneetal2015}.
Specifically,   active systems can  tune   their elastic response  by   modifying  their  effective energy landscape, thereby, developing soft modes endogenously \cite{Sheshkaetal2016,maitravoituriez2020,beheraetal2023,debsankaretal2017,Roychowdhuryetal2022}. In particular, due to the presence of activity,  the   dynamic realizations  of  fragile matter become possible  \cite{Bouchaudetal1997,catesetal1998a,RamaswamyRao2007},  
with marginality successfully  maintained despite the fact that  the overall stability of the system is compromised \cite{CatesEvans2000,Panetal2023,VitelliHecke2012,VitelliHecke2011}.   One may argue that living systems  would be retained in a controlled marginally stable state   if  the   emerging  soft modes facilitate function  \cite{MaoLubensky2018, Lubenskyetal2015, Driscolletal2016, Gagneetal2005,Kleinetal2002,Schweigeretal2007, ZhangMao2017}.

A prototypical context for such behavior is the vast repertoire of rigidities exhibited  by the active cellular cytoskeleton   
 \cite{PollardGoldman2016, WangButlerIngber1993}
whose continuous  reconfiguration  generates  a range of  mechanical  responses \cite{banerjeeetal2020, 
 Matthewsetal2006}. Behind this remarkable  mechanical performance is the   activity  of molecular  motors  that can either stiffen  the cytoskeleton through  actively generated pre-stress  or  fluidize it by facilitating  remodeling \cite{HumphreyKas2002,Koenderinketal2009,Oriolaetal2017,MunroGardel2021,Weirichetal2017,Xietal2019,Ingberetal2014,Chanetal2015}. 
 
A  closely related   manifestation of  the   highly non-conventional  elastic response in such systems  is the  emergence of  extreme  stress and strain concentration which takes the form of  force channeling -- as, for instance,  in actomyosin `stress fibers' \cite{Vignaudetal2021,Lehtimakietal2021} or 
 `dense tethers'  during    active  remodeling of the extracellular matrix \cite{Notbohm2015,Nagel2012,Grekas2021,Wang2014,Chen2019,Ferruzzi2019,Babaei2016,Doha2022,Winer2009}. The emerging localized structures  can  channel forces and can be both assembled and disassembled \cite{Harris1981, Favata2024, Shi2014,Vader2009,Ban2018}.   Ultimately the ability of active systems to generate  highly adaptable  spatial patterns of densification and alignment is behind long distance mechanotransduction indicating the emergence of mechanical pathways which may be as important as the biochemical ones \cite{WangSuo2005,Ingberetal2014,JaaloukLammerding2009,Shivashankar2011,Naetal2008,WangIngber2009,blumenfeld2006}. 

At the microscopic level, the accessibility of the underlying  marginally rigid configurations
 can be attributed in cytoskeleton to buckling of the constituent  semiflexible actin filaments\cite{Roncerayetal2016},  loss of crosslinkers, resulting in their relative sliding   \cite{MurrellGardel2012}, to wrinkling \cite{MullerKierfeld2014}
 and can also be linked to the  stretching-to-bending transition  \cite{Lerner2023,Parvez2023, Buxton2007,Broedersz2014,Salman2019}.
 The underlying  ideas,  allowing cytoskeletal   systems to   maintain   the elastically  extreme   regimes, are  currently   of  considerable interest for   the design   of  artificial biomimetic materials and devices  \cite{Norris2017,Czajkowskietal2022,MiltonCherkaev1995,Almgren1985,Lakes1987,Huangetal2023,Gutierrez1999,BrianeFrancfort2019, Ulrichetal2013,Gomezetal2012,sunetal2012,Mantietal2016}.

In this paper, we identify some of the strategies that enable active solids to reach elastically degenerate regimes characterized by partial rigidity loss. Such regimes have been explored for passive solids \cite{Bouchaudetal1997,catesetal1998a,Grabovskytruskinovsky2013,MaoLubesnky2018,VitelliHecke2011,ChouNelson1996}  and here we demonstrate how active solids, by tuning their internal activity, can navigate the material parameter space in order to  access these states. Additionally, we explore the richness of the  soft modes  emerging in such states by examining their role in supporting  inhomogeneous deformations that can arise in marginal solids subjected to endogenous driving.

 Our analysis of the  elastically  marginal states builds on 
the concept of
\textit{elastic spinodals}, distinguishing them 
from the conventional {\it thermodynamic spinodals}. This   distinction is due to the presence
of long range interactions in elastic systems, arising from the gradient nature of the order parameter and the attendant compatibility constraints
\cite{Karthaetal1995}. While the conventional thermodynamic spinodals are associated with the appearance of
zero eigenvalues of the finite dimensional elastic stiffness
matrix \cite{born1996dynamical}, \textit{elastic spinodals} emerge due to the appearance
of zero eigenvalues of the (infinite dimensional) elastic
differential operator. 

In the context of equilibrium elastic solids, the  idea of distinguishing between elastic and thermodynamic spinodals can be traced to \cite{ChouNelson1996} where  two approaches to elastic stability were 
contrasted: stability with respect to the  components of the displacement field and    stability
with respect to the  components of zero-wave-vector
tensorial strains. To this end,  the Fourier-transformed linearized elastic energy 
was, following  \cite{larkin1969phase},  decomposed into uniform bulk distortions and finite-wave-vector modes. This  approach  was more recently extended to active solids to account for  odd elastic behaviour   \cite{Scheibneretal2020}. 
Here we argue that,   in the case of general nonlinear active solids, the  inhomogeneous soft modes emerging  at the elastic spinodals  can be induced by  internal loading and stabilized   
  in   constrained environments exemplified,  for instance,  by living cells embedded in confluent tissues or attached to patterned substrates \cite{Lehtimakietal2021,Vignaudetal2021}.

 To keep the analysis transparent, 
we deal with the simplest case of  isotropic solids, where we can address in full detail the    accessibility of the  corresponding limits of elastic  stability.   We classify all  finite wave-vector instabilities in such systems linking
   the associated  tensorial stability criterion  with the failure of Legendre–Hadamard conditions and the attendant     ellipticity loss of the equations of elastostatics  \cite{EricksenToupin1956,KnowlesSternberg1978,Grabovskytruskinovsky2013}. 

 The possibility of reaching elastically marginal  states,  which lie beyond the realm of  ordinary materials,  has  been so far  under-emphasized  in the literature.  In this paper, we show that such states  can be realized in active matter.  Furthermore, we demonstrate  that active  systems can potentially cross the \textit{elastic spinodal} thresholds, provided that the  model ensures proper stabilization of these marginal states through nonlinear and higher gradient terms.  
A central focus of this paper is to 
establish that an important  consequence of 
proximity to marginal thresholds is  the  emergence of force channeling along transient structures that can be actively assembled and disassembled.

As part of our analysis, we demonstrate that the {\it macroscopic} active drive toward marginal states is underpinned by a {\it microscopic} stochastic mechanism that destabilizes certain energy minima while generating new ones in the effective free energy landscape.  Specifically, we provide an explicit example of stochastic dynamics in a prototypical system exposed to non-equilibrium noise, which effectively renormalizes linear elastic moduli from positive to negative values.

To summarize, our study highlights the adaptive nature of active matter, where endogenous activity can dynamically alter its elastic response. Such `smart' materials can operate near rigidity thresholds, using a rich spectrum of soft modes to regulate the balance between solidity and fluidity. Moreover, our results suggest that mechanical feedback pathways -- capable of being built up and broken down as needed -- allow active materials to self-drive toward marginal stability, where partial rigidity loss enables the emergence of non-affine 
floppy `mechanisms'.

The paper is organized as follows. In Section 2, we focus on linear elastic isotropic solids and show that, by coupling the activity level  to  the current  state  of passive stress  or strain, the system can renormalize its elastic response in a broad range, potentially reaching both thermodynamic and elastic spinodals. Thermodynamic spinodals are addressed in Section 3 where we discuss the level of elastic degeneracy in such regimes.   Section 4 is dedicated to the study of \textit{elastic spinodals}  where we address the issue of extreme stress concentration in these  regimes and also discuss various   approaches allowing one  to regularize the emerging singularities. Nonlinear active solids are the subject of Section 5 where we consider in full detail a prototypical model extended beyond the limits of linearized marginality. To complement this largely phenomenological approach,  we present  in   Section 6  a microscopic stochastic model  showing how the introduction of a correlated noise can modify  the structure of the energy landscape. Finally,  in Section 7 we summarize the obtained results and present our conclusions.

\section{Linear elastic active solids}
 To elaborate the idea of active accessibility of elastic 
spinodals it is sufficient to consider the simplest case of an
isotropic linear elastic solid in 2D. We start with writing
the total stress $\boldsymbol{\sigma}$  (say, of an active meshwork)  as a sum of
elastic and active terms,
\begin{equation}\label{as}
\boldsymbol{\sigma}= \mathbb{C}^e \boldsymbol{\epsilon}+\boldsymbol{\sigma}^{a}.
\end{equation}
Here,
$
\boldsymbol{\epsilon}=(\nabla\boldsymbol{u}+\nabla\boldsymbol{u}^T)/2  
$ is the   strain tensor and $\boldsymbol{u} (\boldsymbol{x})$ is the displacement field.  The  fourth order elastic  tensor   $\mathbb{C}^e$ has the standard isotropic  form,
\begin{equation}
\mathbb{C}^e_{ijkl}=B^e\,\delta_{ij}\delta_{kl}+\mu^e(\delta_{ik}\delta_{jl}+\delta_{il}\delta_{jk}-\delta_{ij}\delta_{kl}),
\end{equation}
 where $B^e$ and $\mu^e$ are the passive bulk and shear moduli, respectively.  

According to \eqref{as}, in the presence of activity, the total stress $\boldsymbol{\sigma}$ has  an inelastic additive component $\boldsymbol{\sigma}^{a}$ first introduced in  \cite{GmitroScriven1966,FinlaysonScriven1969}.
We consider a special case when the active stress can be represent  in the form 
\begin{equation}
\boldsymbol{\sigma}^a=\mathbb{A}\boldsymbol{M},
\end{equation}
where $\boldsymbol{M}(\boldsymbol{x})$ is a symmetric second order    \textit{fabric} tensor field   characterizing  the density distribution of active agents
and encoding, for instance, a locally diffused dipolar mass
anisotropy of active elements, such as myosin \cite{IrvineShraiman2017, Ioratim-Ubaetal2023,kanatani1984} density distribution. 
The  fourth order tensor $\mathbb{A}$ is again assumed to be  isotropic
\begin{equation}
 \mathbb{A}_{ijkl}=\zeta\,\delta_{ij}\delta_{kl}+\xi\,(\delta_{ik}\delta_{jl}+\delta_{il}\delta_{jk}),
 \end{equation}
where now the coefficients    $\zeta$ and $\xi$   characterize the levels  of spherical and deviatoric activity, respectively.
%

The constitutive turnover of active elements  can be   incorporated  into  the model through the assumption that   the  anisotropy represented by the tensor $\boldsymbol{M}$  
is not arbitrary but is instead enslaved to  the  anisotropy of the current value of stress. It is implied that  the stress state  is  felt by the fabric and that the latter  can be re-orientd in the former. For instance, we assume in this way that  stress anisotropy  can locally  re-orient    cytoskeletal meshworks   \cite{Mirzaetal2023,
 banerjeeetal2020,	Kovacsetal2007, HaganBaskaran2016,Backoucheetal2006}. Here it may be also appropriate to  mention highly relevant studies of  tensegrity models, where load induced stiffening of living cells has been linked to alignment/remodelling of stress fibers along the loading direction \cite{Ingberetal2014}.

\subsection{ Stress induced regulation}
 The simplest quantitative assumption which expresses  this type of stress regulation is  \cite{IrvineShraiman2017, Ioratim-Ubaetal2023}
\begin{equation}\label{M}
\boldsymbol{M}(\boldsymbol{x})=\boldsymbol{M}_0+\mathbb{L}\boldsymbol{\sigma}(\boldsymbol{x}),  
\end{equation}
where $\boldsymbol{M}_0$ is a constant second order tensor and $\mathbb{L}$  is the fourth order tensor  which is again assumed to be isotropic
\begin{equation}
\mathbb{L}_{ijkl}= \bar{L}_b\,\delta_{ij}\delta_{kl}+{L}_s\,(\delta_{ik}\delta_{jl}+\delta_{il}\delta_{jk}),
\end{equation}
now   with  coefficients $\bar{L}_{b}$ and $L_{s}$ representing the level of stress-texture coupling.  In view of the  linearity   assumption in \eqref{M},  we effectively  renormalize the isotropic linear
elastic moduli. 

Indeed, if we   eliminate   $\boldsymbol{M}$ from the stress-strain relation for the total stress  $\boldsymbol{\sigma}$,  we obtain  
\begin{align}
&\boldsymbol{\sigma}=\mathbb{C}^e\boldsymbol{\epsilon}+\boldsymbol{\sigma}^a
=\mathbb{C}^e\boldsymbol{\epsilon}+\mathbb{A}\boldsymbol{M}\nonumber\\
&=\mathbb{C}^e\boldsymbol{\epsilon}+\mathbb{A}(\boldsymbol{M}_0+\mathbb{L}\boldsymbol{\sigma}).
\end{align}
After  algebraic reorganization, this yields  
\begin{equation}\label{sig}
\boldsymbol{\sigma}=(\mathbb{I}-\mathbb{AL})^{-1}\mathbb{C}^e\boldsymbol{\epsilon}+(\mathbb{I}-\mathbb{AL})^{-1}\mathbb{A}\boldsymbol{M}_0,  
\end{equation}
which shows that  
 \begin{equation}\label{sig1}
 \mathbb{C} =(\mathbb{I}-\mathbb{AL})^{-1}\mathbb{C}^e
 \end{equation}
  can be viewed as activity renormalized linear elastic stiffness tensor; the second term in \eqref{sig} plays the role of active prestress.

 Note that in the cytoskeletal setting,  we are essentially postulating in \eqref{M}  that  	the     tensorial kinetic rate   of binding  of active  crosslinkers 
$\mathbf{k}^b=  k^b\,\mathbf{I}  $   is balanced by  a  Bell-type stress dependent  unbinding rate $\mathbf{k}^u\boldsymbol{M}$ with   \cite{Kondaetal2011,Duetal2023,Ranaetal2022,Vernerey2022, Bell1978, Blacketal2011}
 \begin{equation}
 \mathbf{k}^u(\boldsymbol{\sigma})=  k^u   (\mathbf{I}+e^{ \hat{\mathbb{L}}\boldsymbol{\sigma}} ),
 \end{equation}
where $\hat{\mathbb{L}}$   is again a standard isotropic forth order tensor characterized by two constant coefficients. Under these assumptions we obtain that   in the limit of small stress,   $\mathbb{L}=-(k^b/4k^u)\hat{\mathbb{L}}$ and  $\boldsymbol{M}_0 =  (k^b/2k^u)\mathbf{I} $.  

More generally, in the approximation of weak activity, we obtain from \eqref{sig1}
\begin{equation}
\mathbb{C} \approx(\mathbb{I}+\mathbb{AL})\mathbb{C}^e,
\end{equation}
where
\begin{align}
    &(\mathbb{AL})_{ijkl}=\mathbb{A}_{ijpq}\mathbb{L}_{pqkl}\nonumber\\
    &=2\big(\zeta(\bar{L}_b+{L}_s)+\xi {L}_s\big)\delta_{ij}\delta_{kl}
    +2 \xi {L}_s(\delta_{ik}\delta_{jl}+\delta_{il}\delta_{jk})\nonumber\\
    &=:2\zeta\tilde{L}_b\,\delta_{ij}\delta_{kl}+2\xi{L}_s(\delta_{ik}\delta_{jl}+\delta_{il}\delta_{jk}),
\end{align}
and  $\tilde{L}_b:=\bar{L}_b+\Big(1+\frac{\xi}{\zeta}\Big){L}_s$. Furthermore, since
\begin{align}
       &(\mathbb{ALC}^e)_{ijkl}=(\mathbb{AL})_{ijpq}\mathbb{C}^e_{pqkl}\nonumber\\
    &=4\big(\zeta\tilde{L}_b (\lambda^e+\mu^e)+\xi{L}_s\mu^e\big)\delta_{ij}\delta_{kl}\nonumber\\
    &+4\xi{L}_s\mu^e(\delta_{ik}\delta_{jl}+\delta_{il}\delta_{jk}), 
\end{align}
 the renormalized isotropic elastic moduli take the form
\begin{equation}
    B=B^e(1+4\zeta L_b),
\end{equation}
and 
\begin{equation}
    \mu=\mu^e(1+4\xi{L}_s),
\end{equation}
 where 
 $B^e:=\lambda^e+\mu^e$ 
 and 
  $$L_b:=\bar{L}_b+\Big(2+\frac{\xi}{\zeta}\Big(1+\frac{\mu^e}{B^e}\Big)\Big)L_s.$$
Note that  the sign of $L_{b,s}$ determines whether activity leads to a stiffening or softening of the elastic material.
   

\subsection{Strain induced regulation} 
In an alternative  case  when the   fabric  tensor  $\boldsymbol{M}$  is regulated by strain $\boldsymbol{\epsilon}$ (rather than the stress $\boldsymbol{\sigma}$),  we  can similarly assume   that  
 \begin{equation}
\boldsymbol{M}=  \boldsymbol{M}_0+\mathbb{K}\boldsymbol{\epsilon},
\end{equation}
 where the tensor $\mathbb{K}$ is again assumed to be isotropic    with parameters $K_{b,s}$ 
\begin{equation}
    \mathbb{K}_{ijkl}=\bar{K}_b\,\delta_{ij}\delta_{kl}+{K}_s\,(\delta_{ik}\delta_{jl}+\delta_{il}\delta_{jk}).
\end{equation} 
Then the effective ( renormalized) linear elastic response is 
\begin{align}\label{sig2}
&\boldsymbol{\sigma}=\mathbb{C}^e\boldsymbol{\epsilon}+\boldsymbol{\sigma}^a
=\mathbb{C}^e\boldsymbol{\epsilon}+\mathbb{A}\boldsymbol{M}\nonumber\\
&=\mathbb{C}^e\boldsymbol{\epsilon}+\mathbb{A}(\boldsymbol{M}_0+\mathbb{K}\boldsymbol{\epsilon})=(\mathbb{C}^e+\mathbb{AK})\boldsymbol{\epsilon}+\mathbb{A}\boldsymbol{M}_0,
\end{align}
In the low activity level limit the   renormalized  stiffness tensor is  
 \begin{equation}
\mathbb{C}=\mathbb{C}^e+\mathbb{AK},
\end{equation} 
while the second term in \eqref{sig2} again represents active prestress.
To obtain the expressions for the renormalized isotropic linear elastic moduli we write
\begin{align}
    &(\mathbb{AK})_{ijkl}=\mathbb{A}_{ijpq}\mathbb{K}_{pqkl}\nonumber\\
    &=2\big(\zeta(\bar{K}_b+{K}_s)+\xi {K}_s\big)\delta_{ij}\delta_{kl}
    +2 \xi {K}_s(\delta_{ik}\delta_{jl}+\delta_{il}\delta_{jk})\nonumber\\
    &=:2\zeta\tilde{K}_b\,\delta_{ij}\delta_{kl}+2\xi{K}_s(\delta_{ik}\delta_{jl}+\delta_{il}\delta_{jk}),
\end{align}
where $\tilde{K}_b:=\bar{K}_b+\Big(1+\frac{\xi}{\zeta}\Big){K}_s$. The resulting expressions of the effective isotropic elastic moduli are 
\begin{equation}
B=B^e+2\zeta K_b,
\end{equation}
and 
\begin{equation}
\mu=\mu^e+2\xi K_s,
\end{equation}
where again  
 $B^e:=\lambda^e+\mu^e$  and $$K_b:=\tilde{K}_b+\Big(1+\frac{\xi}{\zeta}\Big)K_s.$$  
\begin{figure*}[t]
    \centering
    \includegraphics[scale=0.4 ]{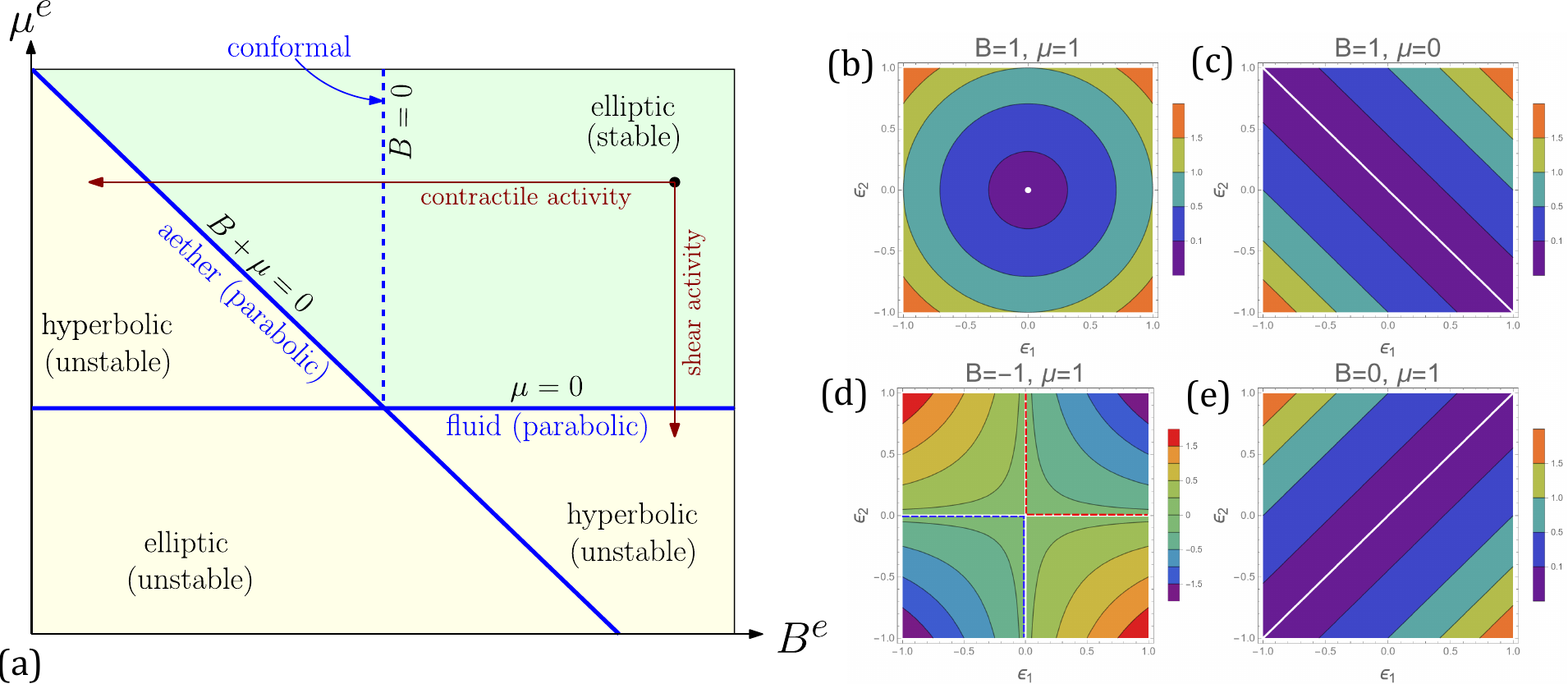}
    \caption{(a) Regime diagram in the space of elastic moduli for  the classical isotropic linear elastic solid.     
    (b-e) Contours of the   linear ``elastic strain energy density'' $w(\epsilon_1,\epsilon_2)$
    for different  $B$ and $\mu$; white line indicates zero `valleys'.    
   In  (b) we see a single strain energy minimum.   It transforms  into a continuous distribution of minima in the thermodynamic spinodal regimes: (c) elastic fluid  and (e) conformal  material.  In (d), we show the two soft mode branches of elastic aether $\text{det}[\boldsymbol{\epsilon}]=0$: no compression aether: $ \text{tr}[\boldsymbol{\epsilon}]>0$ (red dotted line) and no tension aether: $ \text{tr}[\boldsymbol{\epsilon}]<0$ (blue dotted line).}
    \label{Figsi1}
\end{figure*}

Note that, in general, for an active solid, 
the tensors $\mathbb{A}$, $\mathbb{L}$ and $\mathbb{K}$, and hence the renormalized stiffness tensor $\mathbb{C}$, may not exhibit the  major symmetries 
since the stress is not derivable from a free energy. However, in this study, we restrict $\mathbb{C}$ to have the same symmetries of $\mathbb{C}^e$, the passive stiffness tensor, as we deal only with isotropic and shear activity, parametrized by $\zeta$ and $\xi$, respectively.

The elastic  response here  
should be interpreted 
only in an {\it incremental} sense, for instance, as in active vertex models \cite{Farhadifaretal2007}, 
through an infinitesimal work function $dw=\boldsymbol{\sigma}\cdot d\boldsymbol{\epsilon}$.  This exists in general Cauchy elastic materials \cite{Cauchy1828,Truesdell1952,RogersPipkin1963,Ericksen1956,Edelen1977,YavariGorieli2025,Scheibneretal2020}  or   hypoelasticity  \cite{Truesdell1955,Bordigaetal2022}, whose cyclic integral $\oint dw\ne 0$, unless $\boldsymbol{\sigma}=\mathbb{C}\boldsymbol{\epsilon}$ where $\mathbb{C}$ is the effective stiffness tensor with conventional major symmetries.

\subsection{Experimental evidence} 

 As we have seen, in the proposed model the effective moduli  $B$ and $\mu$ can vary in broad limits  characterized by the accessible range of the  
 activity coefficients  $\zeta$  or $\xi$.  Such variations have been indeed recorded in experiments on  living systems.
 
For instance, the implied  stiffening or softening response induced by  endogenous (active)  loading, was observed  in crosslinked actin mesh  \cite{Oriolaetal2017,Chanetal2015,Gardeletal2006,Mizunoetal2007,Chaudhurietal2007,Kovacsetal2007,Koenderinketal2009,Murrelletal2015,Khalilgharibietal2019,Roncerayetal2016}.
In particular, it was shown that  both the effective bulk and shear moduli, $B$ and $\mu$, can stiffen from 1Pa to 100Pa in response to  significant increase in motor activity upto 10-30 fold. 

Initial softening followed by stiffening of crosslinked actomyosin in solution after activating the motors by ATP
addition was observed in \cite{Pegoraroetal2017,AbeMaruyama1971}. It was shown that after ATP addition, the effective shear modulus $\mu$ initially abruptly drops from $\sim$80 Pa to almost $\sim$0 Pa  due to increased
mobility of the elements. At later times, the shear modulus rose   to $\sim$350 Pa as the myosin hydrolyzed the
ATP and the formation of rigor bonds with actin
filaments took place \cite{Pegoraroetal2017,AbeMaruyama1971}. 

To corroborate our assumption of enslavement of the fabric tensor to  the local stress/strain state we refer to {\it in vivo} studies of self-organization of actomyosin cytoskeletal structures in mammalian cells \cite{Bershadsky2017} which showed that
the dissociation of individual non-muscle myosin IIA filaments takes place with rate   $k^u \sim 0.01\,s^{-1}$. On the other hand,
the macroscale formation and reorientation dynamics of actomyosin stress fibers depend on factors at several levels of organization, such as crosslinking density and anisotropy, filament stiffness, substrate anchoring and local adaptive regulation mechanisms, and, hence, it is much slower, taking several minutes to hours \cite{Vignaudetal2021,Lehtimakietal2021}.

Finally, we mention that  in the case of `catch' bonds  we may expect that   $L_{b,s}<0$  or $K_{b,s}<0$  \cite{Kovacsetal2007,debsankaretal2017,Roychowdhuryetal2022}.  This suggests  that the values of the elastic moduli can potentially reach the  elastic  stability  limits specified  in the next Section.   However, to corroborate  such a possibility at a quantitative level, much more extensive experimental studies would be needed. Those are expected to confirm with certainty the proposed time scale separation and   link more securely the increase of endogenous activity in say, actomyosin cytoskeleton, with the attendant changes in the (quasi)elastic properties of the meshwork.



\section{ Thermodynamic spinodals}
 

As we have already mentioned,   in  linearized setting, reaching  thermodynamic spinodal means the loss of the  positive definiteness  of the  effective elastic stiffness tensor $\mathbb{C}$.  For passive solids this condition   is equivalent to the classical condition of the loss of convexity of the corresponding effective linear elastic strain energy density, $w(\boldsymbol{\epsilon})$.
For simplicity, we restrict our attention to  the special case when the actively renormalized effective stiffness tensor $\mathbb{C}$ possesses the same major symmetries as the passive stiffness tensor $\mathbb{C}^e$, and hence, such an effective ``strain energy function'' can still be defined.
For instance, in the simplest nontrivial case of this type, when we deal with a 2D  isotropic linear elastic material,  such actively renormalized ``elastic energy''  (to be interpreted as a work function as discussed in the previous Section) would be  of the form
\begin{equation} \label{SM1}
w(\epsilon_1,\epsilon_2)=B\,(\epsilon_1+\epsilon_2)^2/2+\mu\,(\epsilon_1-\epsilon_2)^2/2,
\end{equation}
where the strain tensor is represented by  the principal strains  $\epsilon_{1,2}:=(\text{tr}\,\boldsymbol{\epsilon}\pm\sqrt{(\text{tr}\,\boldsymbol{\epsilon})^2-4\,\text{det}\,\boldsymbol{\epsilon}})/2$. 
From \eqref{SM1} it is clear that convexity of energy requires positive definiteness of the moduli, $B>0$ and $\mu>0$.  These inequalities  ensure that  the   elastic   body is  stable  independently of the type of the boundary conditions. In this sense, they are sufficient but not necessary for elastic stability. 

We recall  that for isotropic materials  in 2D  the fourth order elastic stiffness tensor $\mathbb{C}$ can be represented as a $3\times 3$ matrix  with eigenvalues $2B$, $2\mu$ and $2\mu$. To identify the (homogeneous) floppy modes activated at the corresponding thermodynamic stability thresholds (thermodynamic spinodals)
 \begin{equation}\label{therm1}
 B=0
  \end{equation}
   and  
   \begin{equation}\label{therm2}
   \mu=0
     \end{equation}
we first observe that at $\mu=0$ there are two degenerate eigenvalues  while at $B=0$ there is one.  It is then easy to see  that at the `elastic fluid' threshold $\mu=0$ two  shear modes soften,  while  at the  `conformal' threshold  $B=0$ (see the explanation for this term below),  a single  dilatation mode becomes floppy.  We can now  illustrate   the  degeneracies of the energy landscape associated with  these  two thermodynamic spinodal regimes. In view of the isotropy, it is again convenient to use the space of principal strains   $ (\epsilon_1,  \epsilon_2)$. Since we are in the linear elastic framework,  the  implied degeneracies always take  the form of  a transformation of a single minimum of the  quadratic ``elastic strain energy density'' at  $ \epsilon_1= \epsilon_2=0 $, see  Fig.  \ref{Figsi1}(a), into the zero energy `valleys'  representing  continuously degenerate minima.   Thus, we obtain a zero dilatation valley  
\begin{equation}
\epsilon_1+\epsilon_2=0
  \end{equation}
   in  the case of  an  elastic fluid which can be then viewed as a `unimodal material'  and a zero  shear valley  
   \begin{equation}
   \epsilon_1-\epsilon_2=0
   \end{equation}
    in the case of  dilatation-insensitive  (conformal) solid which is then a `bimodal material', see Fig.~\ref{Figsi1}(c,e). We use here the language of unimodal and bimodal materials taken from the theory of composite materials \cite{MiltonCherkaev1995}.
  
  
   To illustrate  the emergence of inhomogeneous soft modes  at the thermodynamic spinodals it is sufficient to consider the threshold $B=0$.    As we have seen  in such a limiting regime, purely dilatational  deformations  with  $\epsilon_{xx} -\epsilon_{yy}=0$ and $\epsilon_{xy}=0$ cost no energy. If we  express these conditions  in terms of displacements, we obtain  
  \begin{equation}\label{u1}
   \begin{array}{ll}
   u_{x,x}=u_{y,y}, \\
   u_{x,y}=-u_{y,x}.
   \end{array} 
   \end{equation} 
In view  of \eqref{u1},  the two  components of the displacement field $(u_x,u_y)$ are both harmonic functions.   Moreover,
   \begin{equation}
    \nabla\boldsymbol{u} 
    =\left(
    \begin{array}{cc}
        u_{y,y} & -u_{y,x} \\
        u_{y,x} & u_{y,y}
    \end{array}
    \right)=\mathbf{R}\mathbf{U}
\end{equation}
where $$\mathbf{U}=\sqrt{a^2+b^2}\mathbf{I}$$ is a pure dilatation with  $a=u_{y,y}$, $b=u_{y,x}$, and $\mathbf{R}$ is a rotation by angle $\tan^{-1}(b/a)$. Such  purely dilatational inhomogeneous  soft  displacement modes  are then angle preserving.  It is in view of these features, that the material with $B=0$ is called the conformal material \cite{Czajkowskietal2022,mikhlin73}. 

Note also that the stress in   a  `conformal' solid  is necessarily trace free, 
 \begin{equation}
\text{tr}[\boldsymbol{\sigma}]=0. 
\end{equation} 
This  makes  the  force balance problem  
 \begin{equation}
\text{div}\,\boldsymbol{\sigma}=\mathbf{0}, 
\end{equation} 
statically determinate (isostatic) \cite{broederszetal2011,BallBlumenfeld2002,blumenfeld2004,catesetal1998a,vanhecke2009,blumenfeld2006}.

While the equations of elasticity remain elliptic at the  conformal threshold $B=0$, the elastic body  still becomes unstable in the presence of  an  unconstrained part of the boundary. This is because of the failure of the  `complementing condition' at $B=0$, which marks the effective `loss of ellipticity on the boundary' \cite{SimpsonSpector1985}. This means, for instance,  that for a solid  body with   $B=0$ the traction free boundary is always unstable against surface wrinkling of the form  
\begin{equation}
\boldsymbol{u}(\boldsymbol{x})=Re[\boldsymbol{z}(s)e^{i\mathbf{q}\cdot\boldsymbol{x}}], 
\end{equation}
 where $\mathbf{q}$ is an arbitrary wave vector representing lateral wiggles   while  the function  
\begin{equation}
\boldsymbol{z}(s)=\mu(iq\,\boldsymbol{\nu+\mathbf{q}})e^{-qs} 
\end{equation}
 describes the exponential decay away from the free surface. Here $\boldsymbol{x}$ is a spatial   coordinate on the  (free) surface with normal $\boldsymbol{\nu}$  and  $s$ is the coordinate perpendicular  to this  surface \cite{silhavy-book97}. Since any unconstrained segment of the  boundary can be expected to undergo   a  surface instability as the limit  $B = 0$ is reached,  active materials with free boundaries, e.g., cells in suspension \cite{Maloneyetal2010},  are potentially susceptible to such  instabilities   \cite{sunetal2012, EgorovPalencia2011, silhavy-book97, Lubenskyetal2015}. It is interesting that   despite the presence of these    instabilities, 
 thermodynamic spinodals, corresponding to
both $\mu=0$ and $B=0$, have been successfully achieved  
  by meta-material constructions \cite{Ninarelloetal2022,Hedayati2017, MiltonCherkaev1995,Norris2017,Zhengetal2022, Ulrichetal2013,Almgren1985,Lakes1987,Czajkowskietal2022,sunetal2012,Misseronietal2022,McMahanetal2022,LiKohn2023,Milton2018}.

  
\section{Elastic spinodals}

We associated  the term   \textit{elastic spinodals} with    marginally stable  elastic configurations where   linear stability  is understood  vis-{\`a}-vis the  perturbations with finite wave vector ($\mathbf{q}\ne\mathbf{0}$). To  identify \textit{elastic spinodals} in the simplest setting of 2D isotropic
linear elasticity  we need to  Fourier transform the  Navier's equilibrium equation  
\begin{equation}\label{nav}
\text{div}\,\boldsymbol{\sigma}=
\mu\,\nabla^2\,\boldsymbol{u} + B\,\nabla(\text{div}\,\boldsymbol{u}) =0,
\end{equation} 
which is equivalent to  rewriting it  in the form 
 \begin{equation} \label{A}
\mathbf{Q}(\mathbf{q})\,\hat{\boldsymbol{u}}(\mathbf{q})=\mathbf{0},
\end{equation} 
where
$$
\hat{\boldsymbol{u}}(\mathbf{q})=(1/2\pi)\int_{\mathbf{R}^2}\boldsymbol{u}(\boldsymbol{x})\,e^{-2\pi i\,\boldsymbol{x}\cdot\mathbf{q}} \,d\boldsymbol{x}.
$$ 
In \eqref{A} we introduced the   acoustic tensor \cite{KnowlesSternberg1978,lakes2008negative}
\begin{equation}\label{A11}
    \mathbf{Q}(\mathbf{q})= (B+\mu)\,\mathbf{q}\otimes\mathbf{q} + \mu (\mathbf{I}-\mathbf{q}\otimes\mathbf{q}).
\end{equation}
In terms of  acoustic tensor we can  formulate 
the   conditions  of elastic  marginality 
in the form 
\begin{equation} \label{A1}
\text{det}\,\mathbf{Q}(\mathbf{q})=0.
\end{equation} 
In view of \eqref{A1} marginality is defined here as the condition when  at least one of the 
eigenvalues  of the elasticity operator \eqref{A} is equal to zero.

It is straightforward to see that  the condition \eqref{A1} is satisfied  whenever  either 
  \begin{equation}\label{spin1}
  \mu=0
   \end{equation}
or 
    \begin{equation}\label{spin2}
    B+\mu=0
    \end{equation}
 The  conditions (\ref{spin1}, \ref{spin2})  can be then interpreted as defining the \textit{elastic spinodal} limits which can be also understood as the thresholds  of  marginal stability   \cite{Grabovskytruskinovsky2013,ChouNelson1996,Scheibneretal2020}.

Note that reaching the  thresholds (\ref{spin1}, \ref{spin2})  indicates that our 
Navier's equations \eqref{nav} lose their conventional elliptic nature. As we show below, they also  signal  that  the character  of stress propagation undergoes a fundamental  change  \cite{KnowlesSternberg1978,lakes2008negative,EricksenToupin1956,  
 davilaetal2016,Bordigaetal2021, Tommasietal2012,ZubovRudev2011, Chiritaetal2007}. 
 
 We further observe that the   threshold (\ref{spin1})   corresponds to the state where the   material has just lost  shear resistance  and in what follows we refer to such regimes as ``elastic   liquids''. In contrast, at the second threshold (\ref{spin2}) 
it is the  resistance to longitudinal deformation  that is lost and in what follows, respecting a long tradition ~\cite{EricksenToupin1956,Kelvin1888,Whittaker1910},  we  refer to such regimes as  ``elastic aethers''.

Note further that the  stability limits (\ref{spin1}, \ref{spin2}), which we illustrate in   Fig.~\ref{Figsi1}(d),    are necessarily   weaker than the limits (\ref{therm1}, \ref{therm2})   delineating  thermodynamic  spinodals. We reiterate that the reason is that    the latter deal  only with   homogeneous (affine, finite dimensional) perturbations while the former address a much broader class of inhomogeneous (non-affine, infinite   dimensional) perturbations. In particular, only the thresholds (\ref{spin1}, \ref{spin2})
account for the gradient nature of the order parameter (strain tensor) and the corresponding  elastic compatibility constraints. Therefore thermodynamic spinodals always lie inside 
the \textit{elastic spinodals}.   
As we have already seen, the  elastic regimes located  between the elastic and the thermodynamic spinodals,  where the material  respect  the strong ellipticity conditions   while  violating the thermodynamic stability conditions, are  stable only as long as  their boundary is fully constrained \cite{KnopsPayne1971,silhavy-book97}.  
Therefore an  artificial synthesis of non-biological materials  reaching the   \textit{elastic spinodals}  remains  highly challenging,  even though some design ideas for  spinodal  metamaterial with $B+\mu=0$,  which is    peculiar  due to its `infinitely auxetic' response,  have been already proposed  in the literature, e.g.  \cite{BrianeFrancfort2019}.


 To understand the physical consequences of having an active material at the \textit{elastic  spinodal} limits, we   observe that due    to the scale-free nature of continuum elasticity, at these  thresholds  all wavelengths   become unstable  simultaneously \cite{mikhlin73,  kozhevnikov99}.
  Therefore the ensuing instabilities are massive: in the   elastic liquid regime ($\mu=0$) the  emerging  soft modes are  all solenoidal  fields 
  \begin{equation}
 \text{div}\,\boldsymbol{u}  
=0. 
 \end{equation}
Similarly,  in the case of elastic aether regime ($B+\mu=0$), the emerging soft modes are  comprised of all irrotational fields
    \begin{equation}
    \text{curl}\,\boldsymbol{u} =\mathbf{0}. 
\end{equation}
Note also that the  mechanical response at   \textit{elastic spinodals}  becomes  isostatic (jammed, critical) \cite{blumenfeld2006}. Specifically, in the     elastic  liquid   regime  the   deviatoric  stress must   necessarily vanish 
  \begin{equation}\label{isol}
 \boldsymbol{\sigma}-(1/2)\text{tr}[\boldsymbol{\sigma}]\,\mathbf{I}=\mathbf{0}.
 \end{equation}
Similarly,  in  the  elastic aether  regime  the  determinant of the stress tensor   must necessarily vanish   
    \begin{equation}\label{isoa}
   \text{det}[\boldsymbol{\sigma}]={0}.
   \end{equation}  
We recall that in isostatic states the  system is highly coordinated so  that  both, the nontrivial zero modes  and the states of self-stress,  are only marginally constrained.

\subsection{Green's functions }

  The tendency towards  the formation of  displacement discontinuities  in  \textit{elastic  spinodal} regimes can   be
illustrated most simply by the special structure of the corresponding    2D  Green's
functions. 

\begin{figure}[h!]
\centering
\includegraphics[scale=0.22]{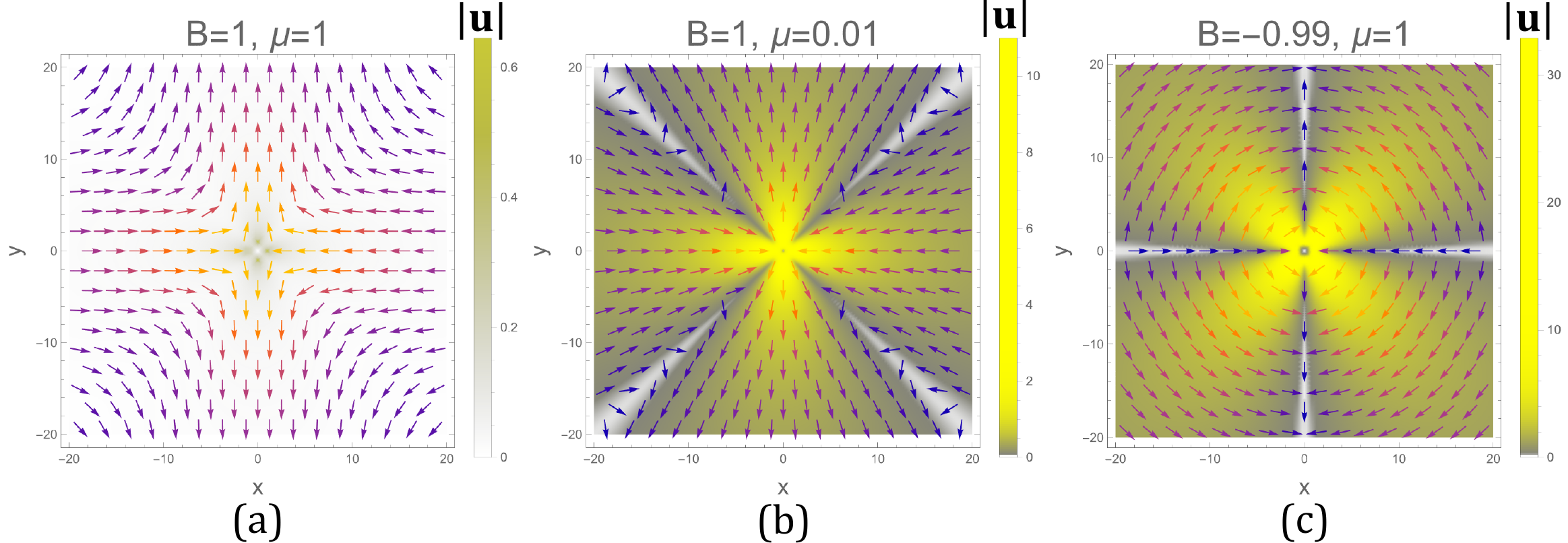}
\caption{Elastic response to a force quadrupole placed in an infinite domain in (a) the stable regime, and in the (b,c) elastic spinodal
regimes.
The displacement field ${\bf u}$, is represented by its direction (arrows) and magnitude (color) in (a) the stable material, (b) close to the  elastic liquid  regime, and   (c)   close to the elastic aether regime, which show the formation of the localized singularities 
of the displacement field ${\bf u}$.
}
\label{fig:elastic111}
\end{figure}

To this end, consider 
the response of an infinite linear elastic   body subject
to concentrated forces.
In the Fourier space the   Green's
function   is just the inverse of the elastic acoustic tensor and in the case of 2D isotropic linear elasticity,
characterized by parameters $B$ and $\mu$, we obtain  \cite{BigoniCapuni2002}
    \begin{eqnarray}
    \hat{G}_{ij}(q_x,q_y)&=&[B\, q_iq_j+\mu\,q_kq_k\,\delta_{ij}]^{-1}\nonumber\\
    &=&\left[
    \begin{array}{cc}
       \frac{\mu q_x^2+(B+\mu)q_y^2}{\mu(B+\mu)(q_x^2+q_y^2)^2}  & -\frac{B q_x q_y}{\mu(B+\mu)(q_x^2+q_y^2)^2} \\
        -\frac{B q_x q_y}{\mu(B+\mu)(q_x^2+q_y^2)^2} & \frac{(B+\mu)q_x^2+\mu q_y^2}{\mu(B+\mu)(q_x^2+q_y^2)^2}
    \end{array}\right].\nonumber\\
    \label{fouriergraansfunction}
\end{eqnarray}
In real space, the obtained result can be illustrated by  presenting  a    displacement  field   generated  by  a pure shear quadrupole at the origin. Suppose that the latter involves a contractile dipole along the x-axis  generated by the point forces $\mathbf{f}^{(1)}=(1,0)$ and $\mathbf{f}^{(2)}=(-1,0)$ which act  at the points $(\mp 1,0)$, respectively,  and an extensile dipole along the y-axis, generated by the point forces $\mathbf{f}^{(3)}=(0,-1)$ and $\mathbf{f}^{(4)}=(0,1)$ which act  at the points $(0,\mp 1)$. The ensuing displacement field takes the form, 
 \begin{eqnarray}
\boldsymbol{u}(x,y)&=&\boldsymbol{u}^{(1)}(x+1,y)+\boldsymbol{u}^{(2)}(x-1,y)+\\
&+&\boldsymbol{u}^{(3)}(x,y+1)+\boldsymbol{u}^{(4)}(x,y-1), 
\end{eqnarray}
 where the corresponding displacement components are  
 \begin{equation}
 {u}^{(\alpha)}_x(x,y)=G_{xx}(x,y)f^{(\alpha)}_x+G_{xy}(x,y)f^{(\alpha)}_y,
 \end{equation}
 and
 \begin{equation}
 {u}^{(\alpha)}_y(x,y)=G_{xy}(x,y)f^{(\alpha)}_x+G_{yy}(x,y)f^{(\alpha)}_y,
 \end{equation}
 where $\alpha\in [1,4]$  and the real space  Green's functions $G_{ij}(x,y)$ are given by  (see, for instance , \cite{Watanabebook}) 
    \begin{eqnarray}
G_{xx}(x,y)&=&\frac{1}{4\pi}\Bigg[-\frac{1}{2}\Big(\frac{1}{\mu}+\frac{1}{B+\mu}\Big)\log (x^2+y^2) \nonumber\\
&&+\Big(\frac{1}{\mu}-\frac{1}{B+\mu}\Big)\frac{x^2}{x^2+y^2}\Bigg],\nonumber\\
 G_{xy}(x,y)&=&\frac{1}{4\pi}\Big(\frac{1}{\mu}-\frac{1}{B+\mu}\Big)\frac{xy}{x^2+y^2},\nonumber\\
 G_{yy}(x,y)&=&\frac{1}{4\pi}\Bigg[-\frac{1}{2}\Big(\frac{1}{\mu}+\frac{1}{B+\mu}\Big)\log (x^2+y^2) \nonumber\\
 &&+\Big(\frac{1}{\mu}-\frac{1}{B+\mu}\Big)\frac{y^2}{x^2+y^2}\Bigg].\label{G}
\label{isotropicgreensfunction}
\end{eqnarray}  
The regular case, when the elasticity equations remain elliptic, is illustrated  in Fig. \ref{fig:elastic111}(a).  The singular   behavior of the functions \eqref{G} 
at \textit{elastic spinodal} limits $B + \mu \to 0$ and $\mu \to 0$ is shown in Fig. \ref{fig:elastic111}(b,c), see  also \cite{Caoetal2018, OuillonSornette2006, RouxHild2002}.
The developing   jump discontinuities, as either the elastic  aether   or  the elastic liquid limits are approached, point towards the unavoidable reduced smoothness of the limiting equilibrium fields.

\subsection{Singular stress fields}
 
To emphasize the ubiquity of stress  concentration phenomena at \textit{elastic  spinodals}, we now  recast the  isostatic conditions \eqref{isol} and \eqref{isoa} in terms of  classical Airy stress function  $\sigma_{xx}=\partial^2_{yy}\chi$, $\sigma_{yy}=\partial^2_{xx}\chi$, $\sigma_{xy}=-\partial^2_{xy}\chi$. 

We recall that in the case of  elastic aether ($B+\mu=0$) the iso-energy lines $\epsilon_1=0$ and $\epsilon_2=0$ in  Fig.\,\ref{Figsi1}(d)  represent the floppy configurations of the reference state. This degeneracy of the energy landscape implies that either of the two principal stresses is zero, so $\sigma_1\sigma_2=0$. This means that  the stress field $\boldsymbol{\sigma}$ is necessarily uniaxial. This property is ultimately behind the isostaticity of this limit suggesting that the equilibrium problem is statically determinate.  We recall that   in this limit the stress distribution  can be indeed found  without invoking any notion of strain field using the  equations 
\begin{equation}
    \text{div}\,\boldsymbol{\sigma}=\mathbf{0},~~~~\text{det}[\boldsymbol{\sigma}]={0}.
\label{stressprobaether}
\end{equation}
Using  Airy stress function we can reduce this system to a \textit{degenerate-elliptic} Monge-Amp\`{e}re equation for $\chi$:
\begin{equation}
    \partial^2_{xx}\chi\,\partial^2_{yy}\chi-(\partial^2_{xy}\chi)^2=0.
    \label{closure:aether}
\end{equation}

The  equation  \eqref{closure:aether}  is known to  admit  (static) shock wave type solutions  with discontinuous gradients of the Airy function  $\chi $ \cite{Caffarellietal1986,angelilloetal2013, Zhengetal2022, cherepanov63}.  In fact,  the  equation \eqref{closure:aether} can be solved geometrically and it is known that  its general solution contains piecewise smooth developable surfaces with zero Gaussian curvature, linked
through folds which can also form corners  with singular Mean and Gaussian curvatures \cite{Caffarellietal1986,Guanetal1999}. The   folds  would then indicate  the locations of stress   channeling   \cite{angelilloetal2013, 
 FraternaliCarpentieri2014, Guanetal1999}. Note that   the same folds are also  indicative of the presence of displacement discontinuities    $\llbracket\boldsymbol{u}\rrbracket \neq 0$ suggesting, for instance,  the emergence of self penetration.

Finally, we stress  that equation \eqref{closure:aether}  \textit{does not} automatically imply the classical strain compatibility condition $$ \textup{curl} \,\textup{curl} \, \boldsymbol{\epsilon}=0,$$ which in terms of Airy function $\chi $ would have been equivalent to an elliptic biharmonic equation $$\nabla^4\chi=0.$$  Therefore the solution of   \eqref{closure:aether} may  be  incompatible  in the sense that  for such solutions,   representing, for instance, an energy-free, `inelastic' self penetration,   a global reference state  would  not exist. 
 


In the case of the elastic liquid ($\mu=0$)  the degeneracy of the reference state leads to the requirement that the deviatoric part of the stress field $\boldsymbol{\sigma}$ is   zero (see the degenerate energy line  $\epsilon_1=-\epsilon_2$ in Fig.\,\ref{Figsi1}(c)). Therefore the stress equilibrium problem, which is again statically determinate,  can be closed as follows
\begin{equation}
    \text{div}\,\boldsymbol{\sigma}=\mathbf{0},~~~~\text{dev}[\boldsymbol{\sigma}]=\mathbf{0}.
    \label{stressprobfluid}
\end{equation}
Indeed, invoking the Airy stress function $\chi$ we obtain the  second order hyperbolic system 
\begin{equation}
    \partial^2_{xx}\chi-\partial^2_{yy}\chi=0,~~~~ \partial^2_{xy}\chi=0.
    \label{closure:fluid}
\end{equation}
The two  scalar equations \eqref{closure:fluid} describe the
  independent shear modes associated with  the strain variables $\epsilon_{xx}-\epsilon_{yy}$ and $\epsilon_{xy}$.  They simultaneously become floppy in our isotropic model    ensuring that in equilibrium necessarily $\sigma_{xx}-\sigma_{yy}=0$ and $\sigma_{xy}=0$, see the second equation in \eqref{stressprobfluid}.

As is well known, in hyperbolic systems boundary conditions specified on
non-characteristic curves `propagate' along the characteristics. 
The first equation in \eqref{closure:fluid} has characteristics $x\pm y=const$; its general solution is therefore of the form $\chi_1(x,y)=f_1(x+y)+g_1(x-y)$. 
The second equation in  \eqref{closure:fluid} has characteristics parallel to the coordinate lines $x=const$ and $y=const$. These characteristics are orthogonal to the family of characteristics of the first equation \eqref{closure:fluid}: the corresponding general solution is of the form the form $\chi_2(x,y)=f_2(x)+g_2(y)$. 
When the the functions $f_{1,2}(x)$ and $g_{1,2}(y)$
are non-smooth, the corresponding surfaces $\chi_{1}(x,y)$ and $\chi_{2}(x,y)$
  will have folds `propagating' along the characteristics.  The corresponding singularities of either Gaussian or mean curvature or both would then indicate the location of singular stress fields.


\begin{figure}[h!]
\centering
\includegraphics[scale=0.12]{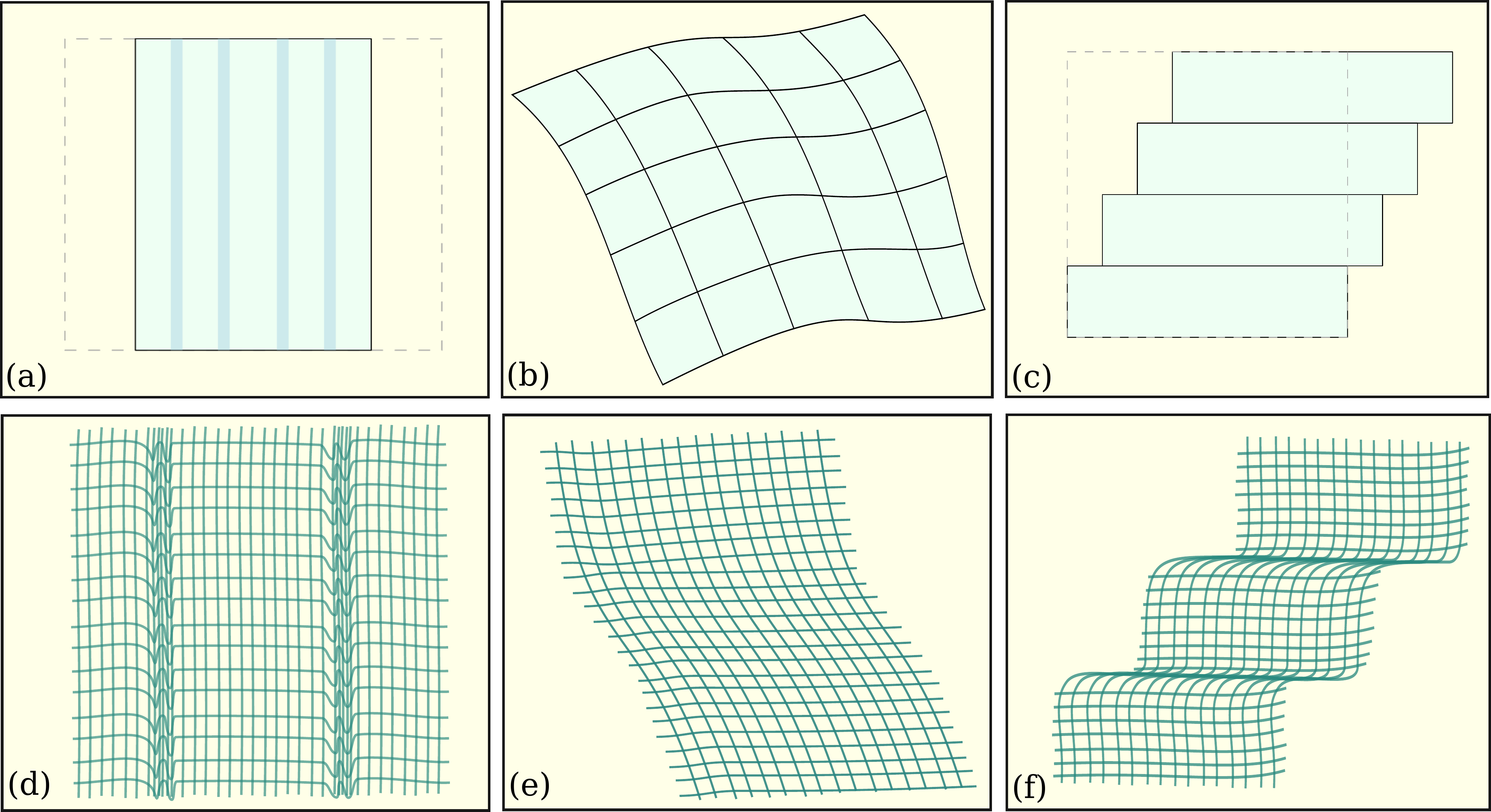}
\caption{
	  Schematic representation of the  singular mechanical response on \textit{elastic spinodals}.  Reference configuration (dashed line) transforms to:   {\it folded} configuration in the elastic aether  regime (a); {\it smooth} configuration in the  elastically stable regime,    $\mu>0, B+\mu>0$ (b);    {\it slipped} configuration in the  elastic liquid regime (c). (d,e,f) Corresponding regularized displacement fields.
}
\label{fig:thermo1}
\end{figure}

 To summarize, in both cases of the elastic aether ($B+\mu=0$) and the elastic  liquid ($\mu=0$)   the system of governing equations  admits  (static) shock-wave-type solutions  with discontinuous gradients  of the Airy function  $\chi(\boldsymbol{x})$, indicating the presence of displacement   jumps   $\llbracket\boldsymbol{u}\rrbracket \neq 0$.   Outside the corresponding  sharp folds   elasticity remains  non-degenerate, the  function $\chi$ remains  smooth and elastic compatibility holds.  
 
 To understand the physical meaning of the discontinuous displacement field we can mention that  in the case of    elastic liquids   the emerging jumps would reflect  a  formation of slip lines, see Fig.\,\ref{fig:thermo1}(c). For elastic liquids   similar  singular solutions  with localized displacement discontinuities are usually discussed in the   dynamic framework  as weak solutions of Euler equations  containing  vortex sheets and other singular structures, e.g.  \cite{Chorin2013,Acharyaetal2017,Chenetal2018,Szekelyhidi2011,LellisSzekelyhidi2022,Brenneretal2016}. Instead,  in the case of   elastic aethers displacement discontinuities  may  describe  either an internal folding of the material, see Fig.\,\ref{fig:thermo1}(a),   or represent an  internal unfolding.  In this sense the elastic aether can perform mechanically as  either  a `no tension' material  \cite{BehringerChakraborty2018,Heyman1966,delpiero1989, Nampoothirietal2020, Blumenfeld2023preprint, angelillo1993,Bouchaud2004} and  a  `no compression' material  \cite{pipkin1986,cherepanov63, Kondoetal1955,Steigmann1990,Panaitescuetal2019}.

\subsection{Fragile nature of elastic spinodals}

As we have already seen, to reach an elastic spinodal state in a body with finite geometry requires  constraints on the boundaries, such as in living cells belonging to confluent tissues, and 
 in cells adhered on micro-patterned substrates~\cite{Vignaudetal2021}. In both cases   the emergence of  system-spanning     `force chains' would depend  on the   anchoring  at the cell boundaries (through focal adhesions \cite{Novak2004} or at adherens junctions \cite{Campasetal2024}).
 Since an arbitrary distribution of the  anchoring sites would  not in general be compatible with the equilibrium structure of the  force chain network,  a dynamic  reorganization  of   the  anchoring configurations  and a concomitant   reconfiguration of the force chains  in the bulk, could be expected until a stable force chain network, compatible with the boundary anchoring distribution, is established  \cite{Bouchaudetal1997}. 
  Such  intermittent mechanical adjustments through building and dismantling of load supporting sub-structures,  has been indeed recorded in  mechanically    fragile   living systems \cite{Floydetal2021, Alencaretal2016, Cardamoneetal2011}.

In the case of elastic aether the  implied   
bulk-boundary correspondence   can be linked to the fact that the effective ``elastic strain energy density''  
\begin{equation}
w(\boldsymbol{\epsilon})=\int_0^{\boldsymbol{\epsilon}}\boldsymbol{\sigma}(\boldsymbol{\epsilon}')\cdot d\boldsymbol{\epsilon}'
\end{equation}
 is  a null-Lagrangian   \cite{lanciaetal1995,verlinde2017}. Indeed, one can see that 
\begin{equation}
\int_{\Omega}w(\boldsymbol{\epsilon})\,dA=\int_{\partial\Omega}\tau_a(\boldsymbol{u},\nabla\boldsymbol{u},\boldsymbol{\nu})\,dx
\end{equation}
where 
\begin{equation}
\tau_a(\boldsymbol{u},\nabla\boldsymbol{u},\boldsymbol{\nu})=-\mu\,e_{ik}e_{jl}\partial_iu_{j}\,u_k\nu_l
\end{equation}
 is a `live load' type surface energy, $e_{ij}$ is the  Levi-Civita symbol, and $\boldsymbol{\nu}$ denotes the unit normal to the boundary.

In the case of elastic liquid, stress is hydrostatic and the equilibrium problem reduces to finding a distribution of scalar pressure in the bulk. The pressure in the bulk is fully controlled by the pressure distribution on the boundary, which must be spatially constant for static situations. Once again we see that  an arbitrary boundary loading  is   not, in general,   compatible with   equilibrium configuration, requiring a dynamic adjustment. More specifically, for  elastic fluids the direct link between the bulk and the surface at equilibrium  can be    illustrated by the straightforward identity   
\begin{equation} \int_\Omega w(\boldsymbol{\epsilon})\,dA \equiv
 \int_\Omega \frac{B}{2}\partial_iu_{i}\,\partial_ju_{j}\,dA =
\int_{\partial\Omega} \tau_f(\boldsymbol{u},\nabla\boldsymbol{u},\boldsymbol{\nu})\,dx,
\end{equation}
where the `live load' type surface energy is now
\begin{equation}
\tau_f(\boldsymbol{u},\nabla\boldsymbol{u},\boldsymbol{\nu})=(B/2)\partial_iu_{i}\,u_j\nu_j.
\end{equation} 
One can see that in equilibrium, when  $\partial_iu_{i}=\text{const} $, the energy depends only on the  imposed change of the total volume while all other types of loading are accommodated dynamically.
 
 We emphasize that such effectively `holographic'   response \cite{Susskind1995},   realized in particular cases through  system spanning stress  singularities,  but  mediated in general by the presence of inhomogeneous (collective) soft modes \cite{Ulrichetal2013,  Gomezetal2012,verlinde2017},     highlights the   fragile nature of the mechanical response   at  \textit{elastic spinodals} \cite{Panetal2023,catesetal1998a,VitelliHecke2012}. It has been argued that in the   case of  cytoskeleton, such fragility  may be supported  by cellular machinery  generating transient tensegrity like networks of `stress fibers' \cite{Vignaudetal2021, Ingberetal2014,KuceraTheryBlanchoin2022}.

 \subsection{ Regularized models} 
 
 In Fig.\ref{fig:thermo1}(a-c)   we illustrated  schematic structure of the generic mechanical response on elastic spinodals, showing separately (a)  the potentially singular response  in  elastic aether  regime, $B + \mu = 0$; (b) the non-singular conventional  regime $\mu > 0$, $B + \mu > 0$; (c) the potentially singular  elastic liquid  regime, $\mu = 0$.  To analyze the mechanical response further, we need to regularize these emergent singularities.
 We note that the singularities appear in  elastic aether  and elastic liquid regimes because the underlying continuum theories suffer from two types of problems:  (i) they are scale free, and (ii) they are elastically degenerate. 
 Therefore, to regularize such singular response one can either bring into the theory an internal (finite) length scale, or compromise the degeneracy.

 In Fig.\,\ref{fig:thermo1}, we provide an illustrative example of the first type of regularization.  Our schematics in the panels  (d-f)  show   the same
characteristic deformation patterns as in  panels (a-c)  but in a regularized model containing a small internal length scale. In such model  an elastic aether regime would be  characterized by small but finite size strain
localization replacing singular   folds.  Elastically non-degenerate  regime would be again described  by smooth maps which would be  basically unchanged vis-{\`a}-vis the non-regularized model. Finally, in the elastic liquid regime  singular slips  would turn into finite size shear bands. The implied   `rounding' of the  displacement discontinuities can be  achieved by switching from local to nonlocal theory,  for instance, by  bringing into the model   higher gradient elasticity (which we discuss in detail later). Another option would be to replace the continuum model by the discrete one.  For comparison of these two approaches to finite length scale regularization, see for instance  \cite{TruskinovskyVainchtein2004,TruskinovskyVainchtein2008}.

 
  We now consider the second  type of regularization  provided by a conventional linear elastic  model which degenerates (saturates)  after small but  finite deformation takes place \cite{Epstein2002}.   To illustrate the idea,  we   present here a simple  example.  

\begin{figure}[h!]
    \centering
    \includegraphics[scale=0.5]{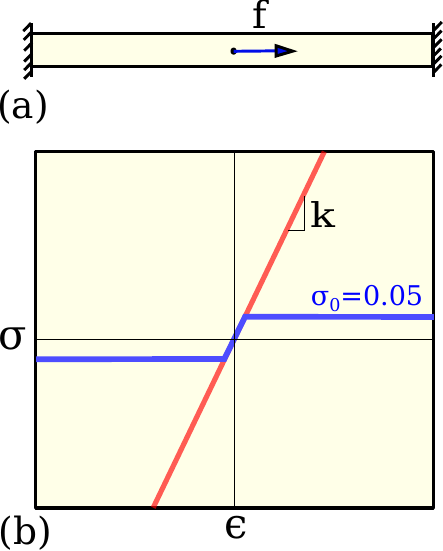}
    \caption{Strain localization in a loaded  active elastic bar: (a) Loading by a body force $f$ at the centre of the clamped bar. (b)  Stress-strain relation of the bar with stiffness $k$ (red line) and stress saturation at $\sigma_0=0.05$ (blue).}
    \label{fig:sea}
\end{figure}

  Consider a 1D elastic bar, fixed at both ends, i.e., with $u(0) = u(L) = 0$, where $u(x)$ is the displacement field and $L$ is the reference length of the bar. Suppose that  such bar is subjected to a concentrated force $f$ at its center, see Fig.\,\ref{fig:sea}(a).  
Suppose further that the material of the bar is `active' in the sense  that a
passive linear elastic response at small strain (characterized by
the modulus $k$), is followed by an activity-induced stress
saturation in both tension and compression. We assume  that the  saturation takes place  at    $\sigma=\pm \sigma_{0}$, see
Fig.\,\ref{fig:sea}(b).

 In this model, 
the fully degenerate \textit{elastic spinodal} limit is reached in the limit    $\sigma_0 \to 0$ when   
non-degenerate elastic range disappears; 
in the opposite limit, $\sigma_0\to\infty$, the material behaves as a passive, purely linear elastic medium with a positive stiffness $k>0$.
We stress that, in contrast to the models with an internal length scale,  the regularization  of a material at the \textit{elastic spinodal} through the introduction of small but finite $\sigma_0$,   does not directly `round'    singularities, with  stress concentration taking place only in the limit of large enough $L$. As we see in the next section, this mechanism of regularization appears naturally in a theory of a nonlinear active solid that includes binding/unbinding kinetics of myosin.



\begin{figure}[t!]
    \centering
    \includegraphics[scale=0.33]{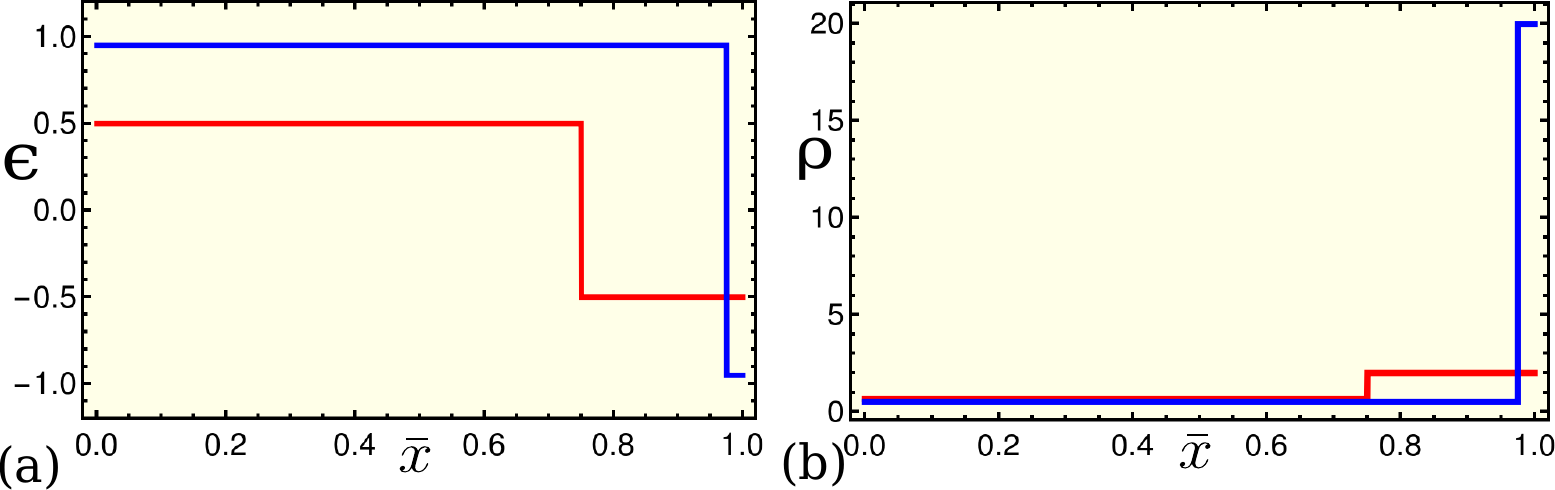}
    \caption{(a) Strain field $\epsilon=du/dx$, corresponding to Fig.\,\ref{fig:sea}, in actual coordinates   $\bar x =x+u(x)$. (b) The deformed material density $\rho(x)=1/ (1+\epsilon(x))$. Parameters  $k=1$, $f=1$, $\sigma_0=0.05$. }
    \label{fig:se}
\end{figure}

Turning back to our example, in the trivial non-degenerate linear elastic limit $\sigma_0\to\infty$ the displacement field is piece-wise linear without showing any signs of localization,  
 \begin{equation}
    u(x)=
     \begin{cases}
      &\frac{f}{2k}x,\,\,\,\,\,\,\,\,0\le x\le \frac{L}{2} \\
        &\frac{f}{2k}(L-x),\,\frac{L}{2}\le x\le L
    \end{cases}\,.
\end{equation}
The corresponding strain field $\epsilon = du/dx$ is then 
\begin{equation}
   \epsilon(x)=
     \begin{cases}
    &\frac{f}{2k},\,\,\, 0\le x\le \frac{L}{2} \\
        &-\frac{f}{2k},\,\frac{L}{2}\le x\le L
    \end{cases}\,.
\end{equation}  
 It is instructive to consider the  density field of the deformed material (assuming that the uniform density of the reference material is 1), defined as
\begin{equation}
\rho(x)=\frac{1}{1+\epsilon(x)},
\end{equation} 
and express it in  the actual (deformed) configuration as $\rho=\rho(\bar{x})$  where 
\begin{equation}
\bar{x}(x)=x+u(x).
\end{equation} 
 Explicitly,
\begin{equation}
   \rho(\bar{x})=
     \begin{cases}
    &\Big(1+\frac{f}{2k}\Big)^{-1},\,\,\, 0\le \bar{x}\le \frac{L}{2}\Big(1+\frac{f}{2k}\Big) \\
        &\Big(1-\frac{f}{2k}\Big)^{-1},\,\frac{L}{2}\Big(1+\frac{f}{2k}\Big)\le \bar{x}\le L
    \end{cases}.
\end{equation}  
We note that the magnitude of the force $f$ must satisfy the strict inequality $f<2k$, so that the deformed length of the right segment of the point of application of the force is non-zero. The corresponding  profiles are illustrated  in Fig.\,\ref{fig:se} (red lines) for $f=k=1$ and $L=1$. 
Consider next the case when the stress threshold is finite $0<\sigma_0<\infty $. Though the resulting equilibrium problem   is nonlinear, it can be solved analytically. In particular, the displacement field becomes
 \begin{equation}
    u(x)=
     \begin{cases}
      &\frac{f-\sigma_0}{k}x,\,\,\,\,\,\,\,\,0\le x\le \frac{L}{2} \\
        &\frac{f-\sigma_0}{k}(L-x),\,\frac{L}{2}\le x\le L
    \end{cases}.
\end{equation}
 The corresponding strain field $\epsilon(x)$ and deformed density field in the actual configuration $\rho(\bar{x})$ are
\begin{equation}
    \epsilon(x)=
     \begin{cases}
      &\frac{f-\sigma_0}{k},\,\,\,\,\,\,\,\,0\le x\le \frac{L}{2} \\
        &-\frac{f-\sigma_0}{k},\,\frac{L}{2}\le x\le L
    \end{cases},
\end{equation}
and
\begin{equation}
    \rho(\bar{x})=
     \begin{cases}
      &\Big(1+\frac{f-\sigma_0}{k}\Big)^{-1},\,\,\,\,\,\,\,\,0\le \bar{x}\le \frac{L}{2}\Big(1+\frac{f-\sigma_0}{k}\Big) \\
        &\Big(1-\frac{f-\sigma_0}{k}\Big)^{-1},\,\frac{L}{2}\Big(1+\frac{f-\sigma_0}{k}\Big)\le \bar{x}\le L
    \end{cases}.
    \label{eq:density}
\end{equation}


Finally, we consider the limit $\sigma_0 \to 0$, when the elastic range around  the reference state $\epsilon=0$ vanishes and the resulting stress strain curve becomes completely flat ({\it elastic spinodal} regime). For $f=k=1$, we note that  in this limit, the length of the segment on the right of the point of application of the force $f$, goes to zero
\begin{equation}
L-\frac{L}{2}\Big(1+\frac{f-\sigma_0}{k}\Big)=\frac{L}{2}\sigma_0\to 0\, .
\end{equation}
However, the density \eqref{eq:density} inside this segment of the bar tends to infinity  
 \begin{equation}
 \rho=\Big(1-\frac{f-\sigma_0}{k}\Big)^{-1}=\sigma_0^{-1}\to\infty\, ,
 \end{equation}
while the density in the rest of the bar remains small  
  \begin{equation}
  \rho=\Big(1+\frac{f-\sigma_0}{k}\Big)^{-1}=(2-\sigma_0)^{-1}\to 1/2\, .
  \end{equation}
To summarize, in the \textit{elastic spinodal} limit $\sigma_0\to 0$, the density profile $\rho(\bar{x})$  localizes around the right boundary. 
Accordingly,    the strains in the left and the right segments, tend  to the limits
\begin{equation}
\pm \frac{f-\sigma_0}{k}=\pm(1-\sigma_0)\to \pm 1\, .
\end{equation}
As a result,  the material of the bar  concentrates   around one point while  forming  an effective rarefied  void-like state everywhere else, see blue line in Fig.\,\ref{fig:se}(b). 

 We interpret this 1D example as  perhaps the most elementary demonstration of the emergence of  `force punctae' in the \textit{elastic spinodal} limit. In 2D at the elastic aether limit,  one can expect a similar tendency towards fraying (fragmentation)  with a homogeneous state turning   into a collection of sparsely distributed dense `force chains', see also \cite{Gangulyetal2017}.

 \section{ Nonlinear elastic active solids}
 

While  the analysis so far has been restricted to linear elastic response, it is imperative to  understand active \textit{elastic spinodals} in a nonlinear setting. This is best realised in the kinetic model, which relates the fields $\boldsymbol{M}$ and  $\boldsymbol{\epsilon}$  via a Bell-type relation. Our main finding is that without assuming weak activity, the renormalized stress-strain response becomes inherently nonlinear and may even turn nonmonotonic.   In this case, the \textit{elastic spinodal} thresholds, isolated in strain space, can be still reached actively,  enabling access to entirely new states disappearing if activity is suppressed.

\subsection{1D Model}

\begin{figure}[h!]
     \centering
     \includegraphics[scale=0.3 ]{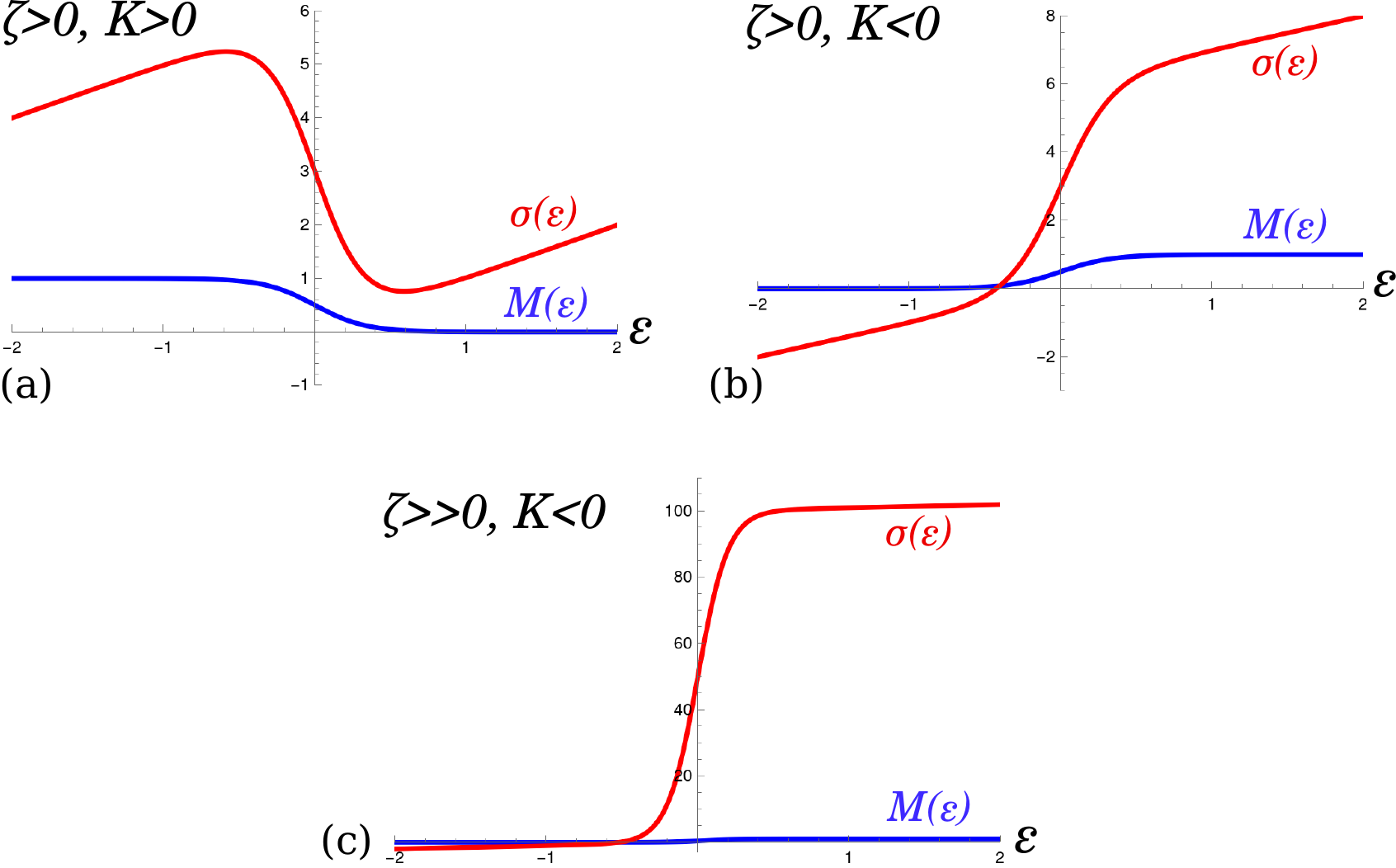}
     \caption{(a) Non-monotonic stress strain behaviour and `catch bond' response at small strains, with saturation at large $|\epsilon|$, of the contractile actomyosin bond for $K>0$. (b) Monotonic stress strain behaviour and `slip bond' response at small strains, with saturation at large $|\epsilon|$, of the contractile actomyosin bond for $K<0$. (c) The slip bond response $K<0$ at large activity $\zeta\gg0$ (here, $\zeta=100$, $K=-10$), tending to the active material with finite elastic range set by a stress threshold $\sigma_0$.}
     \label{fig:sigmarhoform}
 \end{figure}

  To build intuition, we begin with the  simplest    case and consider a 1D  elastic  medium carrying  a scalar fabric-field $M(x, t)$ (as say, associated with myosin density). 
Suppose further that this field is fully enslaved to the  evolving  strain field $\epsilon(x, t)$. 
Under this assumption,  the equation of chemical balance  in the binding-unbinding reaction for the field $M(x, t)$ takes the form,  
\begin{equation}
    k^b-k^u(\epsilon)M=0,\label{activebar}
\end{equation}
where  $k^b$ is the  binding rate which we assume to be strain independent. We take  the unbinding rate $k^u$ to be  strain dependent with a  Bell-form 
\begin{equation}
k^u(\epsilon)=k^u_0(1+e^{K\epsilon}),
\label{bell}
\end{equation}
where $k^u_0$ is a constant. 
 The sign of $K$ determines whether the fabric field exhibits 
a `slip bond' ($K<0$) or a `catch bond' ($K>0$) response \cite{Kovacsetal2007}, see Fig.\,\ref{fig:sigmarhoform}(a,b). 

 The relations \eqref{activebar} and \eqref{bell} give rise to the renormalized nonlinear (quasi)elastic mechanical response, with the following  
stress-strain  relation
\begin{eqnarray}
	 \sigma&=&\mu^e\epsilon+\zeta\,M=\mu^e\epsilon+M_0\zeta(1+e^{K\epsilon})^{-1}\label{effectivestress1}\, .\label{effectivestress1}
\end{eqnarray}
In the `slip bond' case ($K<0$), the stress-strain curve $\sigma(\epsilon)$ remains monotonic with a positive tangent modulus $\sigma'(\epsilon)>0$, ensuring stability (Fig.\,\ref{fig:sigmarhoform}(b)). At high activity $\zeta\gg0$, this behavior resembles (asymptotically) the response adopted in the regularized model discussed previously, where a linear elastic range at small strain terminated in abrupt stress saturation (Fig.\,\ref{fig:sigmarhoform}(c)).
In the `catch bond' case ($K>0$), 
the stress-strain curve $\sigma(\epsilon)$ becomes non-monotonic, with a negative tangent modulus $\sigma'(\epsilon)<0$ near the reference state, leading to instability (Fig.\,\ref{fig:sigmarhoform}(a)). Consequently, 
at sufficiently high activity of catch bond type, the effective non-equilibrium ``elastic strain energy density'' develops a double-well form, destabilizing the homogeneous state $\epsilon=0$. This leads to an elastic decomposition into contracted ($\epsilon<0$) and stretched ($\epsilon>0$) configurations with high and low densities, respectively.

\subsection{Multidimensional  extension }

  In the full tensorial model 
we need to characterize similar  strain
 control, using the   fabric tensor
$\boldsymbol{M}(\boldsymbol{x},t)$.
The corresponding `slaving relation' between $\boldsymbol{M}$ and the strain field $\boldsymbol{\epsilon}(\boldsymbol{x},t)$ takes the form of an algebraic system expressing the conditions of chemical balance:
\begin{equation}
   \mathbf{k}^b-\mathbf{k}^u(\boldsymbol{\epsilon})\,\boldsymbol{M}=\mathbf{0},
    \label{strainmed}
\end{equation}
where 
$\mathbf{k}^b=k^b\mathbf{I}$ is a second order tensor of (strain independent) binding rates and 
$\mathbf{k}^u(\boldsymbol{\epsilon})$ is a second order tensor  of (strain  dependent) unbinding rates.
We assume a generalized  Bell-form for the tensorial unbinding rate: 
\begin{equation}
\mathbf{k}^u(\boldsymbol{\epsilon})=k^u(\mathbf{I}+e^{\mathbb{K}\boldsymbol{\epsilon}}),
\label{Bell2d}
\end{equation} 
where $\mathbb{K}$ is a fourth order tensor encoding the degree of strain dependence of the unbinding kinetics and $k^u>0$ is a constant.  
This leads to the following tensorial  stress-strain relation
\begin{align} \label{renorm}
    &\boldsymbol{\sigma}=\mathbb{C}^e\epsilon+\mathbb{A}\boldsymbol{M}=\mathbb{C}^e\epsilon+\mathbb{A}(\mathbf{k}^{u}(\boldsymbol{\epsilon}))^{-1}\mathbf{k}^b\, .  
\end{align}
The corresponding linearized response  is characterized by the active prestress $$\boldsymbol{\sigma}_{pre}=(k^b/2k^u)\mathbb{A}\mathbf{I},$$ and the  actively renormalized linear elastic stiffness tensor   $$\mathbb{C} =\mathbb{C}^e-(k^b/4k^u)\mathbb{A}\hat{\mathbb{K}}.$$ The remaining 
terms in \eqref{renorm} describes the activity-induced nonlinear  (quasi)elasticity.


\subsection{2D isotropic solid}

%

 To visualize the emergence of new  active energy minima, as the system crosses the {\it elastic spinodal} thresholds,  we now consider in detail  the 2D isotropic case. Using a polynomial approximation of the exponential in \eqref{Bell2d} up to cubic order of strain we can present   the associated effective  ``elastic strain energy density''
     in the form of a Landau-type quartic expansion   
\begin{eqnarray} \label{SM2}
  w(\boldsymbol{\epsilon})&=&(B/2)\epsilon_{mm}\epsilon_{nn}+\mu\,\tilde{\epsilon}_{mn}\tilde{\epsilon}_{mn} \nonumber \\&+&(B'/4)\epsilon_{mm}\epsilon_{nn}\epsilon_{pp}\epsilon_{qq}+\mu'\,\tilde{\epsilon}_{mn}\tilde{\epsilon}_{mn}\tilde{\epsilon}_{pq}\tilde{\epsilon}_{pq}.
  \label{SM2}
\end{eqnarray}
where the linearized strain  
\begin{equation}
\epsilon_{mn}:=(\partial_nu_{m}+\partial_mu_{n})/2
  \end{equation}
   plays the role of a tensorial order parameter. In  \eqref{SM2} we used the standard tensorial notations:   $\epsilon_{mm}$ (summation is assumed over repeated indices) is the   trace  while  $\tilde{\epsilon}_{mn} =\epsilon_{mn}-(\epsilon_{kk})/2\delta_{mn}$ is the  deviatoric part of $\epsilon_{mn}$. 
   Note we  have included in  \eqref{SM2}, both  the   activity renormalized linear elastic moduli $B$ and $\mu$ and the   the activity induced third order moduli   $B'>0$ and $\mu'>0$.  We have for convenience   ignored the cubic terms in   \eqref{SM2}, including a dilation-shear coupling, because the analysis shows that it   has no effect on the discussion of   \textit{elastic spinodals}.

In terms of principal strains the expression \eqref{SM2} can be  rewritten as follows
\begin{equation}
    w(\boldsymbol{\epsilon})= \frac{B}{2}(\epsilon_1+\epsilon_2)^2+\frac{\mu}{2}\,(\epsilon_1-\epsilon_2)^2+\frac{B'}{4}(\epsilon_1+\epsilon_2)^4+\frac{\mu'}{4}(\epsilon_1-\epsilon_2)^4.
    \label{nonlinenergy}
\end{equation}

%
%

 Assuming as before that  
 \begin{equation}
 \hat{\mathbb{K}}_{ijkl}=\hat{K}_b\,\delta_{ij}\delta_{kl}+\hat{K}_s(\delta_{ik}\delta_{jl}+\delta_{il}\delta_{jk})
 \end{equation}
 we can express the  effective isotropic elastic moduli in the form 
\begin{equation}
B=B^e-(\zeta k^b/k^u) K_b,\label{effb}
\end{equation}
\begin{equation}
\mu=\mu^e-(\xi k^b/k^u) \hat{K}_s,\label{effmu}
\end{equation}
where, again,  
\begin{equation}
B^e=\lambda^e+\mu^e
\end{equation}
  and
   \begin{equation}
  K_b=\hat{K}_b+2\Big(1+\frac{\xi}{\zeta}\Big)\hat{K}_s.
  \end{equation}
 To obtain  the explicit expressions for the third order effective moduli  $B'$ and $\mu'$, we ignore the dilation-shear couplings as mentioned before, and write  the cubic term in the small strain expansion for  stress in the form  
\begin{equation}\label{z2}
2(\zeta+\xi)\hat{K}_b^3\,\epsilon^3\,\mathbf{I}+ 8\zeta\,\hat{K}_s^3\,(\text{tr}\,\boldsymbol{\epsilon}^3)\,\mathbf{I}+16\xi\,\hat{K}_s^3\,\boldsymbol{\epsilon}^3.
\end{equation}
From \eqref{z2} we obtain  
\begin{equation}
    B'=(k^b/24 k^u)(\zeta+\xi)\hat{K}_b^3\label{effbp}
\end{equation}
and
\begin{equation}
    \mu'=(k^b/12 k^u)\xi\,\hat{K}_s^3.\label{effmup}
\end{equation}

\subsubsection{Thermodynamic spinodals}

 To locate the thermodynamic spinodals  we need to determine the configurations where the effective ``elastic strain energy density''  \eqref{SM2} loses convexity.  
 Consider an arbitrary  local state with a  strain  $\boldsymbol{\epsilon}=\norm{\epsilon_{mn}}$ and  introduce a small affine strain perturbation characterized by the second order symmetric tensor  $\mathbf{A}=\norm{a_{ij}}$.  The requirement for the tangential stiffness tensor to be positive definite, ensuring the convexity of the  ``elastic energy'', can be formulated in terms of the eigenvalues of the corresponding quadratic form
\begin{eqnarray}
    \mathbb{C}\mathbf{A}\cdot\mathbf{A}&=&B\,(\text{tr}\,\mathbf{A})^2+\mu\,\Big(2|\mathbf{A}|^2-(\text{tr}\,\mathbf{A})^2\Big)\nonumber\\
    &&+3B'\,(\text{tr}\,\boldsymbol{\epsilon})^2(\text{tr}\,\mathbf{A})^2\nonumber\\
    &&+2\mu'\,|\tilde{\boldsymbol{\epsilon}}|^2\Big(2|\mathbf{A}|^2-(\text{tr}\,\mathbf{A})^2\Big)+8\mu'\,(\tilde{\boldsymbol{\epsilon}}\cdot\mathbf{A})^2\, .\nonumber\\
\end{eqnarray}
Here $\mathbb{C}=\norm{C_{ijkl}}$ is  the strain dependent
  stiffness tensor  with components
\begin{eqnarray}
    &&C_{ijkl}(\boldsymbol{\epsilon})= \partial^2w/\partial\epsilon_{ij}\,\partial\epsilon_{kl}=\nonumber\\
    &&B\,\delta_{ij}\delta_{kl}+\mu\,(\delta_{ik}\delta_{jl}+\delta_{il}\delta_{jk}-\delta_{ij}\delta_{kl})+3B'\,(\epsilon_{mm})^2\delta_{ij}\delta_{kl}\nonumber\\
    &&+2\mu'\,\tilde{\epsilon}_{pq}\tilde{\epsilon}_{pq}(\delta_{ik}\delta_{jl}+\delta_{il}\delta_{jk}-\delta_{ij}\delta_{kl})+8\mu'\,\tilde{\epsilon}_{ij}\tilde{\epsilon}_{kl}.
\end{eqnarray}
Choosing $\mathbf{A}$ to be a dilation,  $\mathbf{A}=A\,\mathbf{I}$, we obtain
\begin{eqnarray} \label{SM3}
    \mathbb{C}\mathbf{A}\cdot\mathbf{A}=
  16\Bigg(B+3B'\,(\epsilon_1+\epsilon_2)^2\Bigg)A^2,
\end{eqnarray}
where  $\epsilon_{1,2}$ are again  the  principal strains.  This gives one  condition defining thermodynamic spinodal
\begin{equation}\label{TS1}
B+3B'\,(\epsilon_1+\epsilon_2)^2=0.
\end{equation}
If instead we  choose  $\mathbf{A}$ to be deviatoric,  we get 
\begin{eqnarray} \label{SM4}
 & 0  \le    \mathbb{C}\mathbf{A}\cdot\mathbf{A}
     =  \nonumber \\
    & 2\mu\,|\mathbf{A}|^2+4\mu'\,|\tilde{\boldsymbol{\epsilon}}|^2|\mathbf{A}|^2+8\mu'\,(\tilde{\boldsymbol{\epsilon}}\cdot\mathbf{A})^2 \nonumber \\ &\le  2\Big(\mu+6\mu'\,|\tilde{\boldsymbol{\epsilon}}|^2\Big)|\mathbf{A}|^2\nonumber\\
&\le2\Big(\mu+3\mu'\,(\epsilon_1-\epsilon_2)^2\Big)|\mathbf{A}|^2,
\end{eqnarray}
  where we used the Cauchy-Schwarz inequality 
   $(\tilde{\boldsymbol{\epsilon}}\cdot\mathbf{A})^2\le |\tilde{\boldsymbol{\epsilon}}|^2|\mathbf{A}|^2 $ 
   to obtain the upper bound. 
If the   right hand sides of   \eqref{SM4}  vanishes, the  corresponding strain state $\boldsymbol{\epsilon}$ lies on the thermodynamic spinodal and therefore we obtain the second condition
\begin{equation}\label{TS2}
     \mu+3\mu'\,(\epsilon_1-\epsilon_2)^2=0.
\end{equation}
given  that  dilational  and deviatoric deformations form orthogonal basis in the space of symmetric tensors  which diagonalizes  our  quadratic form,  we can conclude that the two (strain dependent) equations \eqref{TS1} and \eqref{TS2} fully characterize the thermodynamic spinodal in the   model  of 2D isotropic nonlinear elasticity \eqref{SM2}.


\subsubsection{Elastic spinodals} 

We recall that {\it elastic spinodals}   delineate the region in the strain space $\boldsymbol{\epsilon}$ where the strain-dependent acoustic tensor $\mathbf{Q}(\mathbf{q})$ remains positive definite.  In other words, at an  {\it elastic spinodal}, at least one eigenvalue of this tensor vanishes. 

In the material model \eqref{SM2} the expression for the (strain dependent)  acoustic tensor  takes the form
\begin{eqnarray}
    Q_{ik}(\mathbf{q})&=&C_{ijkl}q_j q_l\nonumber\\
    &=&B\,q_i q_k+\mu\,\delta_{ik}+3B'\,(\epsilon_{mm})^2\,q_i q_k\nonumber\\
    &&+2\mu'\,|\tilde{\boldsymbol{\epsilon}}|^2\,\delta_{ik}+8\mu'\,\tilde{\epsilon}_{ij}q_j\,\tilde{\epsilon}_{kl}q_l.
\end{eqnarray}
As in \eqref{A11}, we can use  here  the orthonormal basis $(\mathbf{q}\otimes\mathbf{q},\,\mathbf{q}_{\perp}\otimes\mathbf{q}_{\perp})$, formed by the Fourier space wave vector  $\mathbf{q}$ and its orthogonal complement $\mathbf{q}_{\perp}$,  to rewrite the above expression in the  form
\begin{eqnarray}
    &&\mathbf{Q}(\mathbf{q})=\nonumber\\
    &&\Big(B+3B'\,(\text{tr}\,\boldsymbol{\epsilon})^2+\mu+2\mu'\,|\tilde{\boldsymbol{\epsilon}}|^2+8\mu'\,(\tilde{\boldsymbol{\epsilon}}\mathbf{q}\cdot\mathbf{q})^2\Big)\mathbf{q}\otimes\mathbf{q}\nonumber\\
    &&+\Big(\mu+2\mu'\,|\tilde{\boldsymbol{\epsilon}}|^2+8\mu'\,(\tilde{\boldsymbol{\epsilon}}\mathbf{q}\cdot\mathbf{q}_{\perp})^2\Big)\mathbf{q}_{\perp}\otimes\mathbf{q}_{\perp}.\label{acoustictensor}
\end{eqnarray} 
From this orthogonal representation one can see that vanishing of the eigenvalue associated with the longitudinal modes $\mathbf{q}\otimes\mathbf{q}$ gives   one branch  
of the \textit{elastic spinodal}
\begin{equation}
    B+3B'\,(\text{tr}\,\boldsymbol{\epsilon})^2+\mu+2\mu'\,|\tilde{\boldsymbol{\epsilon}}|^2+8\mu'\,(\tilde{\boldsymbol{\epsilon}}\mathbf{q}\cdot\mathbf{q})^2=0\, ,
    \label{aetherspin}
\end{equation}
which under linearization \eqref{aetherspin} converges to the elastic aether  threshold. Similarly,
vanishing of the eigenvalue associated with the transverse/shear modes $\mathbf{q}_{\perp}\otimes\mathbf{q}_{\perp}$ gives the   other branch  of the \textit{elastic spinodal}
\begin{equation}
    \mu+2\mu'\,|\tilde{\boldsymbol{\epsilon}}|^2+8\mu'\,(\tilde{\boldsymbol{\epsilon}}\mathbf{q}\cdot\mathbf{q}_{\perp})^2=0.
    \label{fluidspin}
\end{equation}
Expectedly, under linearization \eqref{fluidspin} converges to the elastic liquid threshold.
If we now diagonalize the strain tensor using the   basis  $\boldsymbol{\epsilon}= {\epsilon}_{1}\,\mathbf{q}\otimes\mathbf{q}+ {\epsilon}_{2}\,\mathbf{q}_{\perp}\otimes\mathbf{q}_{\perp}$, Eqs.\,\eqref{aetherspin} and \eqref{fluidspin} reduce, respectively, to the following equations defining the segments of the  \textit{elastic spinodal}    in the space of principal strains:
\begin{subequations}
    \begin{align}
        & B+\mu+3B'(\epsilon_1+\epsilon_2)^2+3\mu'\,(\epsilon_1-\epsilon_2)^2=0,\label{aspin}\\
        & \mu+\mu'\,(\epsilon_1-\epsilon_2)^2=0. \label{fspin}
    \end{align}
\end{subequations}
From the expressions of the effective parameters $B$, $\mu$, $B'$ and $\mu'$, given in \eqref{effb}-\eqref{effmup}, we observe that if $B^e,\,\mu^e>0$, and the system exhibits `catch bond' type kinetics ($\hat{K}_{b,s}>0$), increasing activity levels $\zeta,\,\xi>0$  can drive the effective linear elastic moduli $B$ and $\mu$  to change the sign and become negative, while the nonlinear elastic moduli remain positive ($B',\,\mu'>0$). In other words, activity alone can induce the emergence of {\it elastic spinodals}.

\subsubsection{Active remodelling of the ``elastic energy'' landscape} 

\begin{figure}[h!]
\centering
 \includegraphics[scale=0.2]{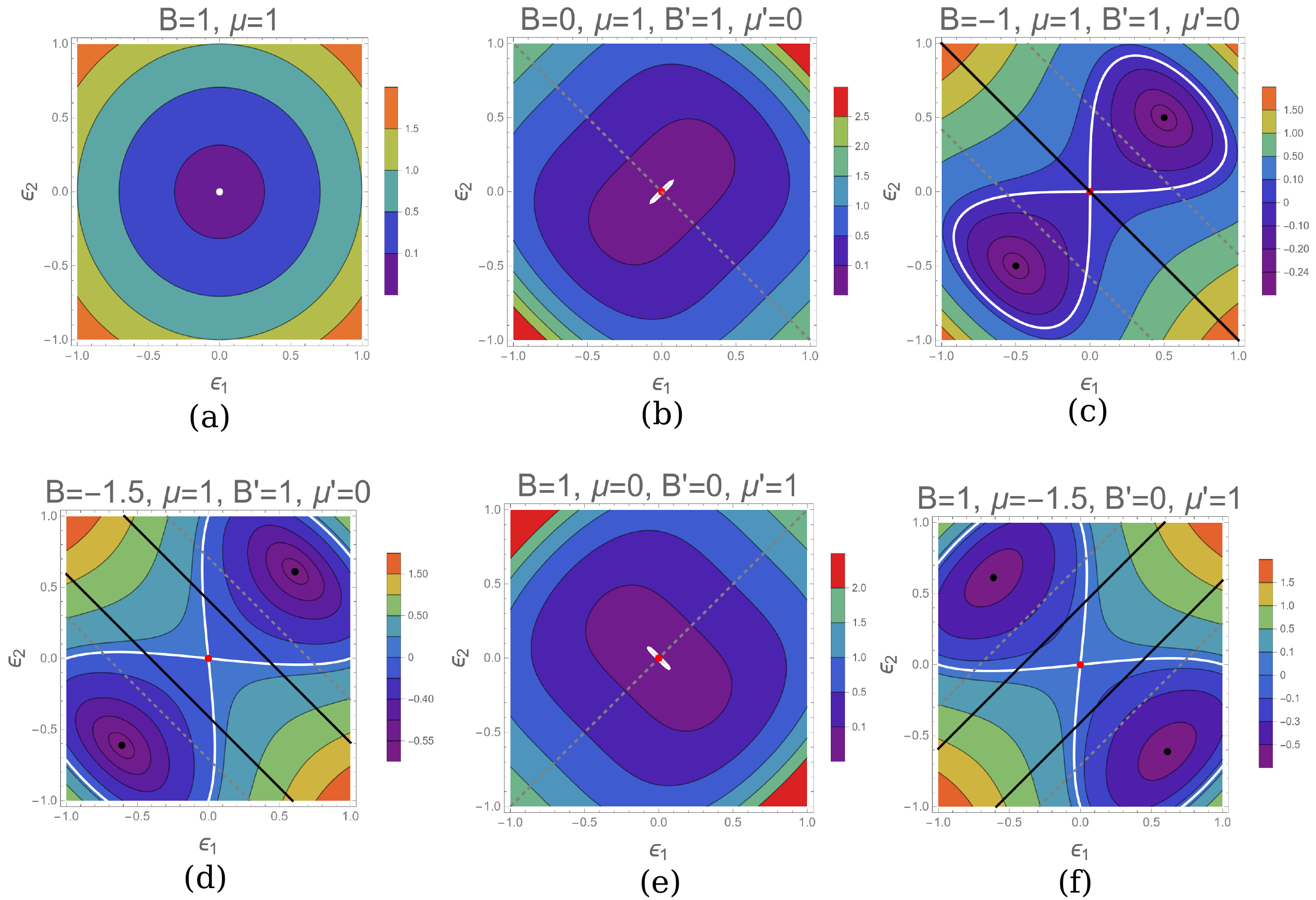}
\caption{Evolution of the effective   ``energy'' landscape  $w(\epsilon_1,\epsilon_2)$  as activities $\zeta,\,\xi$ increase, leading to the appearance of new  minima;
(a) passive solid; active solid at (b) the thermodynamic spinodal, (c) aether, (d) post-aether, (e) elastic liquid, and (f) elastic post-liquid regime. 
In (b-f), elastic and thermodynamic spinodals  are shown by solid and dotted lines, respectively.  
White lines mark zero energy valleys.
}
\label{fig:fluid}
\end{figure}

  As the activity levels $\xi>0$ and $\zeta>0$ change, the  ``elastic energy'' landscapes  (\ref{SM2}, \ref{nonlinenergy})  evolve.  In  Fig.\,\ref{fig:fluid}  we trace  the   modifications in the location of the nonlinear   \textit{elastic spinodals}  in the space of principal strains as the system  crosses  the  `naive' \textit{elastic spinodal} thresholds $\mu=0$ and  $B+\mu=0$ of the linearized model. While the latter  are derived from a linearized analysis of the  ``elastic energy'' around the reference state, 
the `true'  \textit{elastic spinodals}  are generically reached  at a nonzero level of strain. 

Note that in the 
nonlinear regime, the `naive' thresholds 
$\mu=0$ or  $B+\mu=0$ 
mark the points of  activity-induced second order phase transitions (critical points).  Specifically, when the elastic liquid threshold $\mu=0$ is crosses at the activity level $\xi=-\mu^e/(2K_s)$, the two new energy minima emerge along the shear axis, $\epsilon_1-\epsilon_2$, while along the perpendicular hydrostatic axis  $\epsilon_1+\epsilon_2$ the energy remains convex, see 
Fig.\,\ref{fig:fluid}. 
Similarly, passing the elastic aether threshold  $B+\mu=0$ at the activity level $\zeta =-(B^e+\mu^e)/ 2K_b$   generates two new minima along the hydrostatic axis $\epsilon_1+\epsilon_2$   while along the  perpendicular shear axis $\epsilon_1-\epsilon_2$   the convexity of the energy is preserved.  Note also that in the post-liquid regimes with $\mu\lesssim 0$, the two emerging energy minima, shown in Fig.\,\ref{fig:fluid}, are in fact connected and are, therefore, degenerate,  as they represent a continuous family of pure shears related through rigid rotations. The corresponding 
 force balanced mixtures should resemble microstructures encountered in nematic elastomers  \cite{Conti2002}; under some additional assumptions, a pronounced force channeling has been observed numerically in such nonlinear elastic models \cite{Silling1988,Grekas2021}. Instead, in the post-aether regimes with $B\lesssim-\mu$, the emerging energy minima, shown in Fig.\,\ref{fig:fluid}, are  isolated and represent configurations with different densities.  The  
 force balanced microstructures  in such nonlinear elastic materials with purely volumetric phase transitions  are   simple  laminates unless the shear modulus $\mu$ is extremely small   \cite{GrabovskyTruskinovsky2023, GrabovskyTruskinovsky2019}.


 The 
 \textit{elastic spinodals} of  the nonlinear system (black lines in Fig.\,\ref{fig:fluid}) 
 can also be reached by external mechanical loading \cite{Rosakis1990}. 
In the cellular context, this can be achieved by subjecting the body to external forces, say   through embedded magnetic beads,  by using optical tweezers \cite{Ingberetal2014}, or by deforming the substrate \cite{Vignaudetal2021}. External loading  can be also be achieved by the presence of the surrounding cells in a confluent tissue \cite{Ioratim-Ubaetal2023}.
 The same marginal  responses can be also realized in the cellular context actively \cite{Staddon2023,dAvila2016, Qiu1997, 
 KupfermanMamanMoshe2020, Triguero-Platero2023,Fielding2023,Grossman2022}, for instance,  through internal force transducers such as myosin or focal adhesions  \cite{HumphreyKas2002,Kovacsetal2007,Koenderinketal2009}. 
  We further observe that, as in linear case, the thermodynamic spinodals of the nonliner model (gray dotted lines in  Fig.\,\ref{fig:fluid}),   
lie `inside' the elastic spinodals (if one advances from the stable domain).  



 \subsubsection{Force channeling}

\begin{figure}[h!]
    \centering
    \includegraphics[scale=0.31]{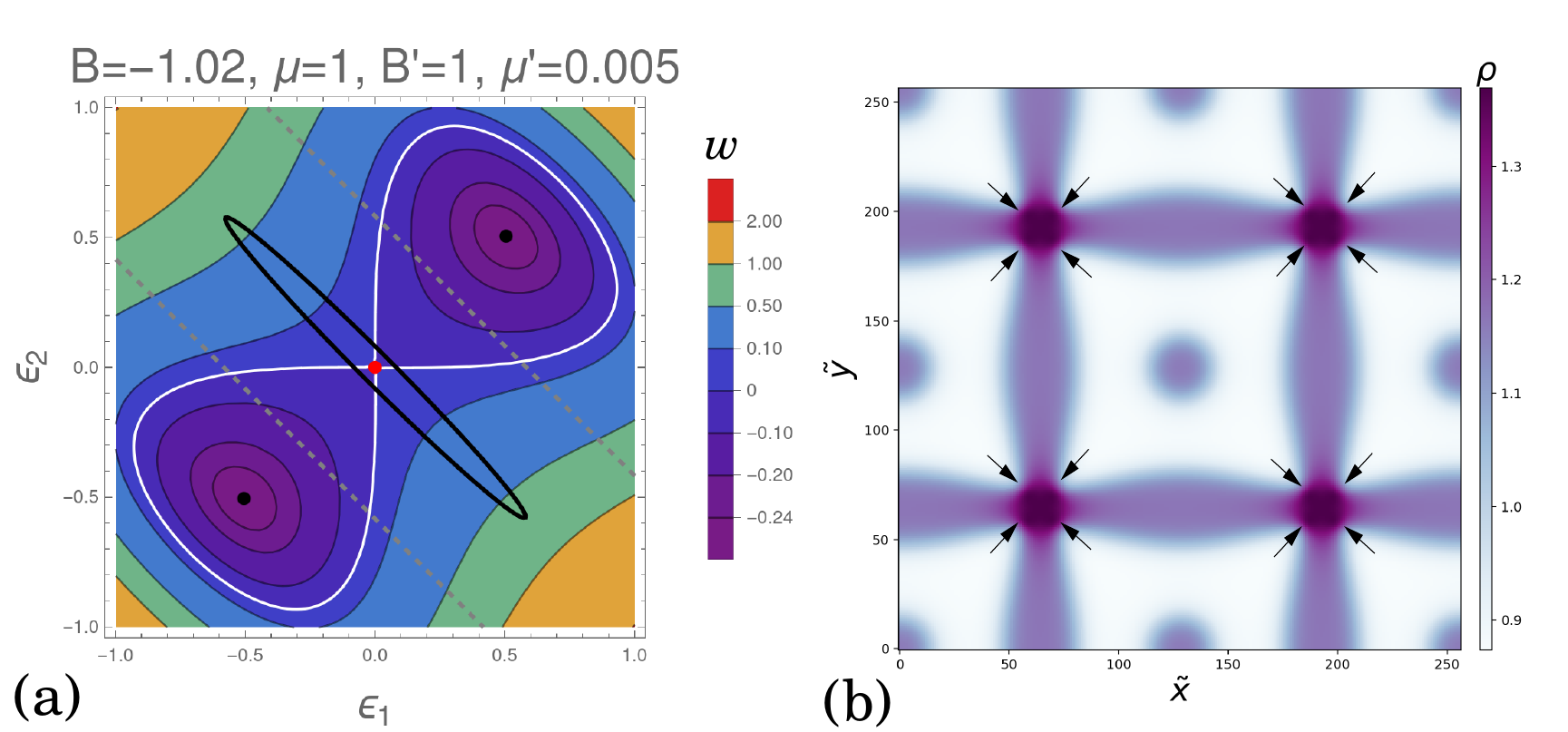}
    \caption{(a) Contour plot of the effective energy landscape in the space of principal strains $w(\epsilon_1,\epsilon_2)$  in a weakly post-aether regime $B=-1.02$, $\mu=1$, $B'=1$, $\mu'=0.005$. Strain localization in such post-aether regime is illustrated in (b), where $\rho$ is the  density of the material, represented in the deformed coordinates.}
    \label{fig:aether1}
\end{figure}

 We now demonstrate the phenomenon of stress channeling in the active material with a (quasi)elastic response governed by \eqref{SM2}. To this end, we choose a slightly post-aether regime characterized by the dimensionless parameters:  $B=-1.02$, $\mu=1$, $B'=1$, $\mu'=0.005$, see Fig.\,\ref{fig:aether1}(b). In this near-critical state, the reference configuration is unstable, and the \textit{elastic spinodal} bounds a spinodal region dominated by small-strain pure shear states $\epsilon_1+\epsilon_2 \approx 0$. 
Such unstable states can be  expected to decompose into a mixture of two active energy wells, corresponding to high- and low-density configurations. In the resulting microstructures  the  force channeling will then  occur along `chain' structures composed of the denser material.

 To `round’ the emerging singularities and control stress concentration, we need to regularize the scale-free model (\ref{SM2}), which will also eliminate mesh dependence in our numerical simulation.  This is achieved by adding a strain gradient term to the work function $w(\boldsymbol{\epsilon})$ \eqref{SM2}, 
\begin{equation} \label{grad}
w(\boldsymbol{\epsilon},\nabla\boldsymbol{\epsilon})=  w(\boldsymbol{\epsilon})+(\kappa/2)|\nabla\boldsymbol{\epsilon}|^2.
  \end{equation}
Here, the parameter $\kappa$ 
introduces an internal length scale into the model, capturing the degree of nonlocal effects. 
To minimize interference of this nonlocality with the formation of microstructure, we chose $\kappa$ to be sufficiently small, $\kappa=0.0001$. 
The expression of  stress  acquires a higher strain gradient correction
\begin{eqnarray} \label{sigma2}
    \boldsymbol{\sigma}(\boldsymbol{\epsilon})&=&\Big[B\,\text{tr}\,\boldsymbol{\epsilon}+B'(\text{tr}\,\boldsymbol{\epsilon})^3\Big]\mathbf{I}+2\mu\tilde{\boldsymbol{\epsilon}}+\mu'\tilde{\boldsymbol{\epsilon}}^3-\kappa\nabla^2\boldsymbol{\epsilon}.\nonumber\\
\end{eqnarray}
  We used \eqref{sigma2} in an overdamped model
\begin{equation} \label{sigma3}
\Gamma\frac{\partial\boldsymbol{u}}{\partial t}=\text{div}\,\boldsymbol{\sigma}+\mathbf{f}, 
\end{equation}
where $\mathbf{f}(\boldsymbol{x})$ is an applied force field and $\Gamma$ is the environmental friction coefficient. In our numerical simulations we used dimensionless variables normalizing  length  with $ (\kappa/\mu)^{-1/2}$, and   time  with $ \Gamma^{-1}(\kappa/\mu)^{-3/2}$. 

 Solving  \eqref{sigma3}   ensures local 
   force balance at each increment  of quasi-static loading. The nature of dynamics changes only when  the system reaches the end of a continuous branch of elastic equilibria.  In such points the equation \eqref{sigma3}  describes   fast (at the time scale of the loading)  mechanical relaxation towards  the closest local force balanced structure.

The system  \eqref{sigma3} was integrated numerically in a 2D square domain with periodic boundary conditions, which allowed us to use a  spectral method  \cite{dedalus2020}. 
To mimic endogenous body forces  we considered the local loading   $\mathbf{f}=\sum_{\alpha=1}^4\mathbf{f}^{(\alpha)}$ representing  two  perpendicular  contractile force dipoles mimicking a center of active contraction.   In view of periodicity of  the boundary conditions we effectively introduced  a periodic lattice  of  quadrupoles, see Fig.\,\ref{fig:aether1}(b). 

As initial conditions we chose  a  homogeneous initial state representing an  unstable reference configuration located inside the narrow spinodal region, Fig.\,\ref{fig:aether1}(a).  A intermediate  outcome of the ensuing  process of elastic spinodal decomposition  is  illustrated  in Fig.\,\ref{fig:aether1}(b) where we show the computed defomred density distribution 
\begin{equation}
\rho(\boldsymbol{x})=\Big(\text{det}\,(\mathbf{I}+\nabla\boldsymbol{u})\Big)^{-1};
\end{equation}
here, we have implicitly assumed that the uniform density of the reference material is 1. For better clarity, $\rho$ is   presented in the coordinates of the deformed configuration 
\begin{equation}
\bar{\boldsymbol{x}}=\boldsymbol{x}+\boldsymbol{u}(\boldsymbol{x}).
\end{equation}
The corresponding  time evolution can
be seen in a movie available at \href{https://drive.google.com/file/d/1Od7nseYqN14xETDUC7cpvzBVW3j2kvCW/view}{SMOV} where we also visualize   the dynamics of dilational $\text{tr}\,\boldsymbol{\sigma}$ and deviatoric $|\tilde{\boldsymbol{\sigma}}|$  stress  measures.

 The density localizations connecting  contractile centers in Fig.~\ref{fig:aether1}(b) and the corresponding depletion outside these regions can be interpreted as  the formation of a  force-chain pattern. Given that our model presents only a minimal prototypical framework, we do not attempt here a detailed comparison with such patterns observed experimentally \cite{Kalaitzidouetal2024, Sopheretal2023, Favata2024}.

 Specifically, a more comprehensive, experimentally calibrated model describe the mechanical properties of active actomyosin cytoskeleton should be tested against observations on cells adhered to micro-patterned substrates \cite{Vignaudetal2021}, confluent tissues \cite{debsankaretal2017}, and in vitro reconstituted systems with controlled mechanical constraints \cite{Krishnaetal2024}. To be realistic, such model should account for the fact that in such active material  the  access to elastically degenerate regimes may arise from micromechanical phenomena such as microbuckling \cite{Roncerayetal2016}, crosslinker loss leading to filament sliding \cite{MurrellGardel2012}, micro-wrinkling \cite{MullerKierfeld2014}, and stretching-to-bending transitions \cite{Lerner2023, Parvez2023, Buxton2007, Broedersz2014, Salman2019}. Capturing these mechanisms requires microscopic modeling, incorporating the dynamic assembly and disassembly of the cytoskeletal elements. This would allow, for instance,  for the description of the observed transient stress fiber formation and their potential merging into force-carrying frames \cite{Ingberetal2014}.

\section{Microscopic stochastic model}


In this section we illustrate the renormalization of elastic rigidity in a simple zero-dimensional stochastic elastic model.
The  analysis below  can be viewed as an elaboration of the model proposed in  \cite{Sheshkaetal2016};  for other related developments see \cite{maitravoituriez2020,beheraetal2023,  
 Driscolletal2016}.

\begin{figure}[t!]
\centering
\includegraphics[scale=0.3]{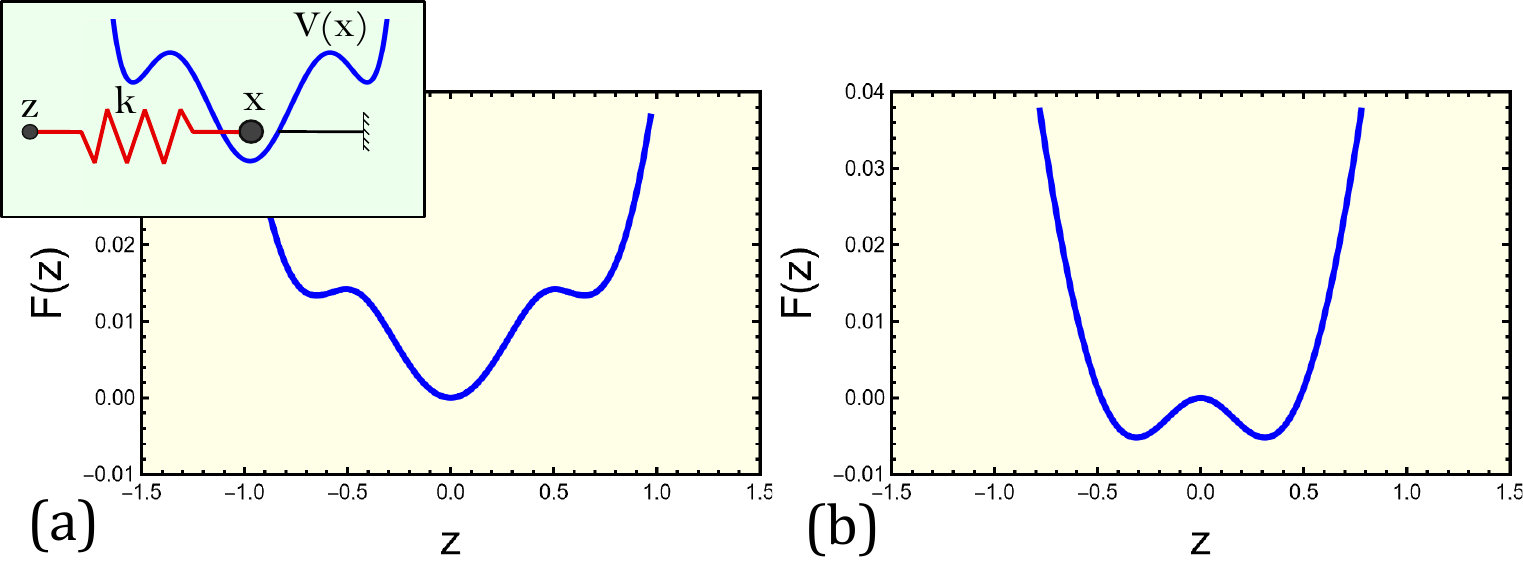}
\caption{
	The effective   potential $F(z)$ at different levels of activity: (a) $A=0$ 
	and (b) $A=0.4$,  with $D=0.01$.}
\label{fig:elastic1111}
\end{figure}

 Consider an  overdamped ratchet-type  stochastic system described by the  Langevin equation and schematically illustrated in the inset in Fig.~\ref{fig:elastic1111}(a) 
\begin{equation}
\dot{x} = -\partial_{x} G + \sqrt{2 D}\eta(t).
\end{equation}
Here  $\eta(t)$  is a  standard white noise with unit variance, $D$ is a measure of temperature  and   
\begin{equation} \label{G1}
G(x,t) = V(x)-xf(t)+k(x-z)^2/2 
\end{equation}
where 
\begin{equation}
V(x)=(1/2)(x^2-0.1)(x^2-0.5)^2 
\end{equation}
is  a polynomial Landau-type   elastic energy (non-renormalized) and $f(t)$ is a time correlated rocking force with zero average. Note that  we have also implicitly assumed that  the configuration  of this non-equilibrium system is continuously probed  through a spring with  stiffness $k$.  In this case the variable $z$ plays the role of an external control parameter, see the inset in Fig.~\ref{fig:elastic1111}(a).

The     effective  force  exerted on the spring  and recorded by the  external loading device  can be found by  averaging  the response $x(t)$  over   ensemble and over time,  which gives 
\begin{equation}
T (z)= k[z-
\underset{t\rightarrow\infty}{\lim}(1/t)\int_{0}^t \int_{-\infty}^{\infty} x p(x,t') dx dt']. 
\end{equation}
%
To find the   probability distribution $p(x,t)$ one must  find the corresponding time dependent solution of   the Fokker-Planck equation 
\begin{equation}
\partial_{t}p= \partial_{x}\left[
p \partial_{x}G+D \partial_{x}p\right]. 
\end{equation} 
After the function $T (z)$ is known, one can determine the   non-equilibrium (quasi)elastic energy simply by integration
\begin{equation}
F(z)=\int^z T(s)ds.
\end{equation} 
The function $F(z)$  would then play  the role of an active  renormalization of the   original passive elastic energy. 

Suppose further that the correlated driving signal $f(t)$ is  periodic  and or analytical simplicity  choose  it the simplest piece-wise constant form
\begin{equation}
f(t)=A(-1)^{n(t)},
\end{equation}
where  $n(t)=\lfloor  2t/\tau \rfloor$.  The advantage of this choice is that the ensuing mathematical problem can be solved analytically in the adiabatic approximation, see  \cite{Magnasco1993} for a similar analysis.

To develop such an  approximation, we assume that the time scale $\tau$ is large comparing to the time scale of the thermally induced barrier crossing in the original potential $V(x)$. We can then focus on the time intervals where the driving force is constant
 \begin{equation}
f(t)\equiv A.
\end{equation}
 The solution of
the corresponding time-independent Fokker-Planck equation
\begin{equation}
\partial_{x}\left[p \partial_{x}G+D \partial_{x}p\right]=0
\end{equation}
can be written explicitly   \begin{equation}
p_{A}(x)=Z^{-1}e^{-G(x,z)/D},
\end{equation} 
where we introduced the  normalization constant 
\begin{equation}
Z=\int_{-\infty}^{\infty}e^{-G(x,z)/D}dx.
\end{equation}  
Given  our  assumption of time scale separation,   we can now write the time-averaged probability distribution in the form
\begin{equation}
p_{\text{ad}}(x)=(p_{A}(x)+p_{-A}(x))/2.
\end{equation}

Our results are illustrated in Fig.\,\ref{fig:elastic1111}(a,b) where we show the evolution of the elastic energy of the system as the level of activity measured by the rocking amplitude $A$ changes from zero, $A=0$,   to  a finite value   $A>0$. Note that the energy minimum at $z=0$ which corresponds to stable equilibrium in the passive system  is transformed by activity   into an unstable state representing a local energy maximum. More specifically,  under the  growing activity level (increasing $A$), the energy minimum at $z=0$  first flattens at  a critical value  $A=A_c$, which marks  the state  of zero rigidity, and then disappears signaling a second order phase transition.  The subsequent growth of activity level reconfigures the energy landscape further favoring  two  symmetric  actively-supported ground states. 

While the proposed model is obviously oversimplistic, it shows the possibility of active modification of energy landscapes due to exposure of mechanical systems to nonequilibrium reservoirs. The model can be made more comprehensive if we consider a  network of such  active springs while  representing activity not only through correlated noise but also through hysteretic delays and non-reciprocity \cite{SheshkaTruskinovsky2014}. 
The resulting model would serve as a microscopic description  behind  the emerging fragility at the coarse grained continuum scale.

\section{ Conclusions}

  In this paper we investigated mechanisms of rigidity loss in isotropic elastic solids and explored how active materials can be directed toward marginal and fragile regimes. A key contribution of this work is refining in the active matter context of the concept of \textit{elastic spinodals}, while stressing their fundamental difference from the more conventional thermodynamic spinodals. While the emergence  of degenerate acoustic modes at finite wave numbers has been explored before \cite{ChouNelson1996, Scheibneretal2020}, the active realization of materials where such modes can become mechanically operative has been underplayed. It has been also overlooked that at \textit{elastic spinodal} thresholds, inhomogeneous soft modes not only soften the overall rigidity but also shift the  modality of stress propagation from diffusion-like to force channeling.

Unlike thermodynamic spinodals, \textit{elastic spinodals} emerge due to the long-range interactions inherent in elastic systems, where strain — serving as the order parameter — has a gradient structure. In nonlinear elasticity, such states  describe minimally stable, marginal deformation gradients. While the conventional engineering designs prioritize instead maximally stable equilibria, there is a growing evidence that biological systems operate near marginal stability which allows them to exploit  a  repository of zero energy modes. Moreover, as we show, active systems can self-tune towards such states and then  exploit them for functional advantage.

Limiting attention only to the simplest case of elastic marginality in isotropic solids, we were able to explore the explicit mechanisms of   actively  reaching the  mechanical regimes with  partial rigidity loss. The most striking manifestation of such regimes  is  stress localization, leading to force channeling along transient low-dimensional substructures, such as stress tethers and force chains. Our key conjecture is that in the presence of nonlinear adaptive feedback mechanisms  these structures can be actively assembled and disassembled. In particular, we argued that cellular cytoskeleton exemplifies an active  material which is self-tuned towards  marginal regime where it  dynamically balances solidity and fluidity.

 Finally, our study suggests that artificial active materials, equipped with appropriate feedback control mechanisms, can be designed to operate at marginal stability, leveraging complex non-affine soft modes. While such behavior is absent in passive materials, the potential for extreme mechanical responses could inspire bio-mimetic implementations using existing metamaterial approaches \cite{WuZiaetal2023, Huangetal2023, DykstraCoulais2023, ZhengGuoetal2021, WangYangetal2023, SinhaMukhopadhyay2023, ZhangSunetal2022, ChenYaoetal2023, Veenstraetal2015}. On the theoretical side, extending these results to anisotropic \cite{BigoniGourgiotis2016, Epstein2002, EverstinePipkin1971} and  more general  Cauchy  elastic materials would likely reveal new fragile soft modes and spinodal regimes. Additionally, future studies should investigate fragility in active solids with elastic incompatibility, where activity may drive defect proliferation and quasi-plastic behavior \cite{Tangetal2024}.

\section{ Acknowledgments} 
The authors thank A. Gupta and G. Zurlo for helpful discussions. AR and MR acknowledge support from the Department of Atomic Energy (India), under project no.\,RTI4006, and the Simons Foundation (Grant No.\,287975), and computational facilities at NCBS. MR acknowledges a JC Bose Fellowship from DST-SERB (India). LT was supported by the grants ANR-17-CE08-0047-02, ANR–21-CE08-MESOCRYSP and ERC-H2020-MSCA-RISE-2020-101008140.

\end{document}